\documentclass[12pt]{article}
\usepackage{bm}
\usepackage{amsmath}
        \numberwithin{equation}{section}
\usepackage{amssymb}
\usepackage[dvips]{graphicx}
\usepackage{bbold}
\usepackage{multirow}
\usepackage{enumerate}
\usepackage{ulem}
\usepackage{color}

\usepackage{hyperref}

\usepackage[utf8]{inputenc}
\begin{document}

\begin{titlepage}
\begin{flushright}
TIT/HEP-660 \\
May,  2017
\end{flushright}
\vspace{0.5cm}
\begin{center}
{\Large \bf
Quantum  periods 
and
prepotential 
in 
${\cal N}=2$ SU(2) SQCD 
}
\lineskip .75em
\vskip 2.5cm
{\large  Katsushi Ito, Shoichi Kanno and Takafumi Okubo }
\vskip 2.5em
 {\normalsize\it Department of Physics,\\
Tokyo Institute of Technology\\
Tokyo, 152-8551, Japan}
\vskip 3.0em
\end{center}
\begin{abstract}
We study  ${\cal N}=2$ SU(2) supersymmetric QCD with massive hypermultiplets deformed in the  Nekrasov-Shatashvili limit of the Omega-background. 
The prepotential of the low-energy effective theory is determined by  the WKB solution of the 
quantum  Seiberg-Witten curve.
We calculate the 
deformed Seiberg-Witten periods  around the massless monoplole point  explicitly up to the fourth order in the deformation parameter.  
\end{abstract}
\end{titlepage}
\baselineskip=0.7cm

\section{Introduction}
 The Seiberg-Witten (SW) solution \cite{Seiberg:1994rs,Seiberg:1994aj}
 of the prepotential
 of ${\cal N}=2$ supersymmetric gauge theory
 enables us to understand both weak and strong coupling physics of the theory such as instanton effects,
 the duality of the BPS spectrum \cite{Seiberg:1994rs,Seiberg:1994aj}
and nonlocal superconformal fixed point \cite{Argyres:1995jj,Argyres:1995xn}.
In the weak coupling region,  the Nekrasov partition function 
\cite{Nekrasov:2002qd,Nekrasov:2003rj}, where the
 gauge theory is defined in the $\Omega$-background \cite{Moore:1997dj}, provides an exact formula of the prepotential including the
 nonperturbative instanton effects. The Nekrasov partition function can be computed with the help of the localization technique.
 At strong coupling region, however, we do not know the localization method to reproduce the prepotential around the massless monopole point.

 The Nekrasov function is  related to the conformal block of two dimensional conformal field theory \cite{Alday:2009aq,Gaiotto:2009ma} and also the partition function of topological string theory \cite{Huang:2009md}.
The analysis of the conformal block with insertion of the surface operator
\cite{Alday:2009fs,Maruyoshi:2010iu,Awata:2010bz}  leads to the concept of
 the quantum Seiberg-Witten  curve. The solution of the quantum curve 
gives
 the low-energy effective theory of the $\Omega$-deformed theories, which are 
 parametrized by two deformation parameters
$\epsilon_1$ and $\epsilon_2$.
In the Nekrasov-Shatashvili limit \cite{Nekrasov:2009rc}  
of the $\Omega$-background, where one of the deformation parameters $\epsilon_2$
is set to be zero, the quantum curve becomes the ordinary differential equation.
The quantum SW curve is obtained from the quantization procedure 
of the symplectic structure defined by the SW differential
\cite{Poghossian:2010pn} where the parameter $\epsilon_1$ plays a role of the Planck constant $\hbar$.
In particular, the SW curve for SU(2) Yang-Mills theory becomes the Schr\"odinger equation with the sine-Gordon potential and the higer order corrections to the deformed period integrals in the weak coupling have been calculated by using the WKB analysis \cite{Mironov:2009uv}.
This was generalized to ${\cal N}=2$ $SU(N)$ SQCD \cite{Zenkevich:2011zx}.
Note that the SW curve for $\mathcal{N}=2^*$ $SU(2)$ gauge theory corresponds to the Lam\'e equation and the deformed period integrals also have been calculated by using the WKB analysis \cite{Beccaria:2016wop,He:2016khf}.
One can derive the Bohr-Sommerfeld quantization conditions which are nothing but the Baxter's T-Q relations of the integrable system \cite{Zenkevich:2011zx,Mironov:2009dv,Popolitov:2010bz}.
The deformed period integral agrees with that obtained from the Nekrasov partition function. 

It is interesting to study perturbative and 
 non-pertubative quantum corrections in the strong coupling region of the moduli space, which might change the strong coupling dynamics of the theory.
In \cite{He:2010xa}, the perturbative corrections around the massless monopole point in the ${\cal N}=2$ SU(2) super Yang-Mills theory have been studied.
In \cite{Krefl:2014nfa}, the 1-instanton correction in $\hbar$ to the dual prepotential has been calculated.
In \cite{Basar:2015xna,Kashani-Poor:2015pca,Ashok:2016yxz,Basar:2017hpr}, the non-perturbative aspects of the $\hbar $ expansion in $\mathcal{N}=2$ theories have been studied.
The purpose of this work is to study systematically perturbative corrections in $\hbar$ to the prepotential
at strong coupling where the BPS monopole becomes massless  for ${\cal N}=2$ SU(2) SQCD
with $N_f=1,2,3,4$ hypermultiplets. We investigate quantum corrections to the
period integrals of the SW differential and the prepotential up to the fourth order in the deformation parameter $\hbar$.

This paper is organized as follows:
In Section 2, we review the quantization of the SW curve and the quantum periods for ${\cal N}=2$ SU(2) SQCD. 
In Section 3, we show that the quantum correction can be expressed by acting the differential operator on the undeformed SW periods in detail.
In Section 4, we calculate the quantum periods in the weak coupling region for $\mathcal{N}=2$ $SU(2)$ SQCD and confirm that they agree with those obtained from the Nekrasov partition function.
In Section 5, we study the expansions of the periods around the massless monopole point in the moduli space. We consider how the effective coupling and the massless monopole point are deformed by $\hbar$. 
In Section 6, we add some comments and discussions.

\section{Quantum SW curve for  ${\cal N}=2$ SU(2) SQCD}
The Seiberg-Witten curve for ${\cal N}=2$ $SU(2)$ gauge theory with $N_f$
($=0,\ldots,4$) 
hypermultiplets is given by
\begin{align} \label{SW_curve}
 K(p)-\frac{\bar{\Lambda} }{2} (K_+(p)e^{ix}+K_-(p)e^{-ix})=0, 
\end{align}
where 
$\bar \Lambda =\Lambda_{N_f}^{2-\frac{N_f}{2}}$ with $\Lambda_{N_f}$ being a QCD scale parameter for $N_f\leq 3$ and $\bar{\Lambda}=\sqrt{q}$ for $N_f=4$. 
Here $q=e^{ 2\pi i \tau_{UV}}$ and $\tau_{UV}$ denotes the UV coupling constant \cite{Dorey:1996bn,Alday:2009aq}.
$K(p)$ and $K_{\pm}(p)$ are defined by
\begin{align} \label{coulomb_moduli}
 K(p)=
 \left\{
 \begin{array}{cc}
 p^2-u,  & N_f=0,1\\
 p^2-u+{\Lambda_2^2\over8},  & N_f=2
 \\
 p^2-u +\frac{\Lambda_3}{4}(p+{m_1+m_2+m_3\over2}), & N_f=3
 \\
 (1+{q\over2})p^2-u +{q \over4}p\sum_{i=1}^4 m_i +{q\over8}\sum_{i<j}m_i m_j , & N_f=4
 \end{array}
 \right.
\end{align}
and
\begin{align} \label{higgs_moduli}
 K_+(p)=\prod ^{N_+}_{j=1}(p+m_j) &,&
 K_-(p)=\prod_{j=N_++1}^{N_f}(p+m_j), 
\end{align}
where $u$ is the Coulomb moduli parameter and $m_1,\ldots, m_{N_f}$ are mass parameters.
$N_+$ is a fixed integer satisfying $1\leq N_+\leq N_f$.
The curve (\ref{SW_curve}) can be written into the standard form \cite{Hanany:1995na}
\begin{align}
y^2=K(p)^2-\bar{\Lambda}^2 K_+(p)K_-(p)
\label{eq:curve2}
\end{align}
by introducing $y=\bar{\Lambda}K_+(p)e^{ix}-K(p)$.
The SW differential is defined by
\begin{align}
\lambda=pd\log {K_-\over K_+} -2 \pi i p dx.
\end{align}
Let $\alpha$ and $\beta$ be a pair of canonical one-cycles on the
curve. 
The SW periods are defined by
\begin{align} \label{SW_periods}
a=\int_{\alpha}p(x) dx, \quad
a_D=\int_{\beta}p(x) dx,
\end{align}
where $p(x)$ is a solution of (\ref{SW_curve}).
Then the prepotential ${\cal F}(a)$ is determined by
\begin{align}
a_D={\partial {\cal F}(a)\over \partial a}.
\end{align}

The SW differential defines a symplectic form $d\lambda_{SW}=dp\wedge dx$
on the $(p,x)$ space. The quantum  SW curve is obtained by regarding the
coordinate $p$ as the differential operator $-i \hbar {d\over dx}$.
We have the differential equations
\begin{align}
\left(
 K(-i\hbar \partial_x))-\frac{\bar \Lambda }{2} (
e^{{ix\over2}}K_+(-i\hbar \partial_x)e^{{ix\over2}}+e^{-{ix\over2}}K_-(-i\hbar \partial_x)e^{-{ix\over2}}
\right)\Psi(x)=0,
\label{eq:qswc1}
\end{align}
where $\partial_x={\partial\over \partial x}$.
Here we take the ordering prescription of the differential operators as in \cite{Zenkevich:2011zx}.
This differential equation is also obtained by observing
the relation between the quantum integrable models and
the SW theory in the Nekrasov-Shatashvili (NS) limit of the
$\Omega$-background \cite{Mironov:2009uv}.
The same differential equation is also obtained from the insertion of
the degenerate primary field corresponding to the
 surface operator in the two-dimensional conformal field theory
 \cite{Alday:2009fs,Maruyoshi:2010iu,Awata:2010bz}.
 
 In this paper, we will choose $N_+$ such that the differential equation becomes the second order differential equation of the form:
 \begin{align}
(\partial_x^2 +f(x)\partial_x+g(x))\Psi(x)=0.
 \end{align}
Then we convert this equation into the Schr\"odinger type equation by introducing $\Psi(x)=\exp(-{1\over2}\int f(x)dx)\psi(x)$:
\begin{align} 
(-\hbar^2\partial_x^2+Q(x))\psi(x)=0,
\label{eq:qsw1}
\end{align}
where
$Q(x)=-{1\over \hbar^2}(-{1\over2}\partial_x f-{1\over4}f^2+g)$.
In the  case of  SU(2) SQCD, it is found that 
$Q(x)$ is  expanded in $\hbar$ as
\begin{align}
Q(x)=Q_0(x)+\hbar^2 Q_2(x).
\end{align}

The quantum SW periods are defined by the WKB solution of the
equation (\ref{eq:qsw1}):
\begin{align}
\psi(x)=\exp\left(
{i\over \hbar}\int^x P(y)dy
\right), 
\end{align}
where
\begin{align}
 P(y)=\sum_{n=0}^{\infty}\hbar^n p_n(y)
  \label{eq:wkb1}
\end{align}
and $p_0(y)=p(y)$.
Substituting the expansion (\ref{eq:wkb1}) into (\ref{eq:qsw1}), we have
the recursion relations for $p_n(x)$'s. 
Note that  $p_{n}(x)$ for odd $n$ becomes a total derivative and 
only  $p_{2n}(x)$ contributes the period integral.
The first three $p_{2n}$'s are given by 
\begin{align} \label{recursion_relation}
  p_0(x)&=i \sqrt{Q_0},\\
  p_2(x)&={i\over2}{Q_2\over \sqrt{Q_0}}+{i\over 48}{\partial_x ^2 Q_0\over Q_0^{3\over2}}, \label{recursion_relation2}\\
  p_4(x)&=-{7i\over 1536}{(\partial_x^2 Q_0{})^2\over Q_0^{7\over2}}+{i\over 768}{\partial_x ^4Q_0\over Q_0^{5\over2}}-{i Q_2 \partial_x^2 Q_0\over 32 Q_0^{5\over2}}+{i\partial_x^2 Q_2\over 48 Q_0^{3\over2}}-{i Q_2^2\over 8 Q_0^{3\over2}}, \label{recursion_relation4}
\end{align} 
up to total derivatives.
Then the quantum period integral $\Pi=\int P(x) dx=(a, a_D)$ along the cycles $\alpha$ and $\beta$ can be expanded in $\hbar$ as
\begin{align}
  \Pi=\Pi^{(0)}+\hbar^2 \Pi^{(2)}+\hbar^4 \Pi^{(4)}+\cdots, 
\end{align}
where $\Pi^{(2n)}:=\int p_{2n}(x)dx$.

Now we study the equations satisfied by the quantum SW periods.
It has been shown that the undeformed (or classical) SW periods $\Pi^{(0)}$ obey the third order differential equation with respect to the moduli parameter $u$ called the 
Picard-Fuchs equation \cite{Ceresole:1994fr,Klemm:1995wp,Ito:1995ga,Ohta:1996hq,Ohta:1996fr,Masuda:1996xj}. 
Note that $\partial_u p_0$ is the holomorphic diffrential on the  curve.
When we  write the curve (\ref{eq:curve2}) in the form 
\begin{align}
y^2=\prod_{i=1}^{4}(x-e_i),
\label{eq:curve3}
\end{align}
where the weak coupling limit corresponds to  $e_2\rightarrow e_3$ and $e_1\rightarrow e_4$,
we can evaluate the periods
\begin{align}
\partial_u \Pi^{(0)}=\int \partial_u p_0 dx=\int{dp\over y}
\end{align}
by the hypergeometric function. 
Then  by using quadratic and cubic transformations \cite{Erdelyi, Masuda:1996xj}, one  finds that in the weak coupling region, where $u$ is large,  the classical periods $\partial_u a^{(0)}$ and $\partial_u a^{(0)}_D$ are given by
\begin{align} 
\partial_u a&={\sqrt{2}\over2}(-D)^{-1/4}F\left( {1\over12},{5\over12};1; z\right),\label{general;a}\\
\partial_u a_D&=i\frac{\sqrt{2}}{2} (-D)^{-\frac{1}{4}} \biggl[\frac{3}{2\pi} \ln 12 F\left( \frac{1}{12}, \frac{5}{12}; 1; z \right) -\frac{1}{2\pi} F_{*} \left( \frac{1}{12}, \frac{5}{12}; 1; z \right) \biggr], \label{general;aD}
\end{align}
where $z=-{27\Delta\over 4D^3}$
and the weak coupling region corresponds to $z=0$. 
Here
$\Delta$ and
$D$ for the curve (\ref{eq:curve3}) are defined by
\begin{align}
\Delta&=\prod_{i<j}(e_i-e_j)^2, \label{general;discriminant}\\
D&=\sum_{i<j}e_i^2 e_j^2-6\prod_{i=1}^4 e_i-\sum_{i<j<k}(e_i^2e_j e_k+e_i e_j^2 e_k+e_i e_j e_k^2). \label{general;D}
\end{align}
$\Delta$ is the discriminant of the curve.
$F(\alpha,\beta;\gamma;z)$ and $F_*(\alpha,\beta;\gamma;z)$ are 
the hypergeometric functions
defined by
\begin{align}
\begin{split}
F(\alpha , \beta ;\gamma ;z ) &=\sum _{n=0} ^\infty \frac{(\alpha )_n (\beta )_ n}{n ! (\gamma )_n } z^n ,\\
F_*(\alpha , \beta ;1;z) &=F(\alpha , \beta ;1 ;z )\ln z+\sum _{n=0} ^\infty \frac{(\alpha )_n (\beta )_ n}{(n !)^2}  \sum_{r=0} ^{n-1} \biggl( \frac{1}{\alpha +r } +\frac{1}{\beta +r} -\frac{2}{1+r} \biggr) z^n .
\end{split}
\end{align}
Changing the variable from $z$ to $u$, the hypergeometric differential equation for $F\left( {1\over12},{5\over12};1; z\right)$ leads to the Picard-Fuchs equation for $\frac{\partial \Pi^{(0)}}{\partial u}$.
It takes the form
\begin{align}
\frac{\partial^3\Pi^{(0)}}{\partial u^3}+p_1 \frac{\partial^2 \Pi^{(0)}}{\partial u^2}+p_2 \frac{\partial \Pi^{(0)}}{\partial u}=0,
\label{eq:pf1}
\end{align}
where $p_1$ and $p_2$ are given by
\begin{align}
p_1&=
{\partial_u (-D)^{1/4} \over (-D)^{1/4}} -{\partial_u ^2 z\over \partial_u  z}
+{\gamma-(1 +\alpha+\beta)z\over z(1-z)}\partial_u z, \label{general;p1}\\
p_2&={\partial_u^2 (-D)^{1/4} \over (-D)^{1/4}}+
{\partial_u (-D)^{1/4} \over (-D)^{1/4}}
\Bigl\{
-{\partial_u ^2z\over \partial_u z}
+{\gamma-(1 +\alpha+\beta)z\over z(1-z)}\partial_u z
\Bigr\}
-{\alpha\beta \over z(1-z)} \left(\partial_u z \right)^2
\end{align}
with $\alpha={1\over12}$, $\beta={5\over12}$ and $\gamma=1$. 
For the SW curve  (\ref{SW_curve}) with $N_f\leq 3$, the Picard-Fuchs equations (\ref{eq:pf1}) agree with those in \cite{Ohta:1996hq,Ohta:1996fr}.
Note that for massless case,  the Picard-Fuchs equation turns  out to be  the second order differential equation for $\Pi^{(0)}$ \cite{Ito:1995ga}.

The higher order correction $\Pi^{(k)}$ to the SW period $\Pi^{(0)}$ is determined by acting a differential operator $\hat{\cal O}_k$ on $\Pi^{(0)}$ \cite{Huang:2009md,Mironov:2009dv,He:2010xa,Huang:2012kn}:
\begin{align}
\Pi^{(k)}=\hat{\cal O}_k \Pi^{(0)}.
\label{eq:hodo1}
\end{align}
There are various ways to represent the differential operator $\hat{\cal O}_k$. For example, one can use the first and second order differential operators with respect to $u$ to express $\Pi^{(k)}$ as
\begin{align} \label{general;relation}
\Pi^{(k)}=\left(X_k^1\frac{\partial^2}{\partial u^2} +X_k^2 \frac{\partial}{\partial u} \right)\Pi^{(0)}.
\end{align}

% N_f=0 case
Let us study the simplest example, the $N_f=0$ theory. 
We have the quantum SW curve (\ref{eq:qsw1}) with the sine-Gordon potential:
\begin{align}
Q(x)=-u-{\Lambda_0^2 \over2}(e^{ix}+e^{-ix}).
\end{align}
The SW periods $\Pi^{(0)}$ satisfy the Picard-Fuchs equation \cite{Ceresole:1994fr}:
\begin{align}
\frac{\partial^2 \Pi^{(0)}}{\partial u^2}-{1\over 4(\Lambda_0^4-u^2)}\Pi^{(0)}=0.
\label{eq:pfnf0}
\end{align}
The discriminant $\Delta$ and $D$ are given by
\begin{align}
\begin{split} \label{Nf=0;discriminant}
\Delta =256 \Lambda _0^8 \left(u^2-\Lambda_0^4\right), \qquad D=12 \Lambda_0^4-16 u^2.
\end{split}
\end{align}
The second and fourth order quantum corrections are given by \cite{Huang:2009md,Mironov:2009uv,He:2010xa}
\begin{align}
\Pi^{(2)}&=\left({1\over12}u \frac{\partial^2}{\partial u^2}+{1\over24}\frac{\partial}{\partial u}\right)\Pi^{(0)}, \label{Nf=0;periods2}\\
\Pi^{(4)}&=\left(\frac{75 \Lambda_0^8-4 u^4+153 \Lambda_0^4 u^2}{5760 \left(u^2-\Lambda_0^4\right){}^2}\frac{\partial^2}{\partial u^2} -\frac{u^3-15 \Lambda_0^4 u}{2880 \left(u^2-\Lambda_0^4\right){}^2}\frac{\partial}{\partial u}\right) \Pi^{(0)}.
\label{eq:nf0pi2pi4}
\end{align}
With the help of the Picard-Fuchs equation (\ref{eq:pfnf0}), we find a simpler formula for $\Pi^{(4)}$:
\begin{align}
\Pi^{(4)}&=\left({7\over 1440}u^2\frac{\partial^4}{\partial u^4}+{1\over 48}u\frac{\partial^3}{\partial u^3}
+{5\over 384}\frac{\partial^2}{\partial u^2}\right)\Pi^{(0)}.
\label{eq:nf0pi4}
\end{align}

In the weak  coupling region where $u\gg \Lambda_0^2$, substituting (\ref{Nf=0;discriminant}) into (\ref{general;a}) and (\ref{general;aD}), we can obtain $a^{(0)}$ and $a_D^{(0)}$ by expanding (\ref{general;a}) and (\ref{general;aD}) around $u=\infty $ and integrating with respect to $u$.
The quantum SW periods can be obtained by applying (\ref{Nf=0;periods2}) and (\ref{eq:nf0pi4}) on $a(u)$ and $a_D(u)$:
\begin{align}
\begin{split}
a(u)=&\left( \sqrt{\frac{u}{2}} -\frac{\Lambda_0}{16 \sqrt{2}}\left(\frac{\Lambda_0^2}{u}\right)^{3/2}+\cdots \right)+\frac{\hbar^2}{\Lambda_0} \left(-\frac{1}{64 \sqrt{2}}\left(\frac{\Lambda_0^2}{u}\right)^{5/2}-\frac{35}{2048 \sqrt{2}}\left(\frac{\Lambda_0^2}{u}\right)^{9/2} +\cdots \right) \\
&+ \frac{\hbar^4}{\Lambda_0^3}\left(-\frac{1}{256 \sqrt{2}}\left(\frac{\Lambda_0^2}{u}\right)^{7/2} -\frac{273}{16384 \sqrt{2}}\left(\frac{\Lambda_0^2}{u}\right)^{11/2} +\cdots\right) +\cdots, \\
a_D(u)=&-\frac{i}{2\sqrt{2}\pi} \left[-4\sqrt{2} a(u) \log \frac{8u}{\Lambda_0^2} +\left(8 \sqrt{u}-\frac{\Lambda_0^4}{4 u^{3/2}} +\cdots  \right) + \frac{\hbar^2}{\Lambda_0} \left( -\frac{1}{6 \sqrt{u}}-\frac{13}{96}\left(\frac{\Lambda_0^2}{u}\right)^{5/2} +\cdots \right)\right. \\
&\left.
 +\frac{\hbar^4}{\Lambda_0^3} \left( \frac{1}{720 u^{3/2}}-\frac{63}{1280} \left(\frac{\Lambda_0^2}{u}\right)^{7/2} +\cdots \right) +\cdots \right],
\end{split}
\end{align}
up to the fourth order in $\hbar$.
It has been checked that the quantum curve
reproduces the prepotential obtained from the NS limit of the Nekrasov
partition
function \cite{Mironov:2009uv,He:2010xa}. 

We can also consider the quantum SW periods in the strong coupling region.
For example, at $u=\pm \Lambda_0^2$
where monopole/dyon becomes massless,  by solving
the Picard-Fuchs equation in terms of hypergeometric function, we can compute the SW periods \cite{Klemm:1995wp}.
For the computation of the deformed SW periods, it is convenient to use 
(\ref{eq:nf0pi4}) rather than (\ref{eq:nf0pi2pi4}) since the coefficients in (\ref{eq:nf0pi2pi4}) become singular at $u=\Lambda_0^2$.
We then  find the  expansion of the SW periods around $u=\Lambda_0^2$, which are given by \cite{He:2010xa}
\begin{align}
\begin{split}
a_D(\tilde u) =& i\left( \frac{ \tilde{u}}{2 \Lambda_0}-\frac{ \tilde{u}^2}{32 \Lambda_0^3}+\cdots \right) +\frac{i\hbar^2}{\Lambda_0} \left( \frac{1}{64}-\frac{5}{1024}\left( \frac{\tilde u}{\Lambda_0^2} \right)+\cdots \right) \\
&+ \frac{i\hbar^4}{\Lambda_0^3} \left(-\frac{17}{65536}+ \frac{721}{2097152}\left( \frac{\tilde u}{\Lambda_0^2} \right)+\cdots \right)+\cdots,\\
a(\tilde u)=&\frac{i}{2\pi}\left[ a_D(\tilde u) \log \frac{\tilde u}{2^5 \Lambda_0^2} +i\left(- \frac{\tilde{u}}{2 \Lambda_0}-\frac{3 \tilde{u}^2}{64 \Lambda_0^3}+\cdots \right)+ \frac{i\hbar^2}{\Lambda_0} \left(\frac{1}{24}\left( \frac{\tilde u}{\Lambda_0^2} \right)^{-1}+\frac{5}{192}+\cdots  \right) \right. \\
&\left. +\frac{i\hbar^4}{\Lambda_0^3} \left(\frac{7}{1440}\left( \frac{\tilde u}{\Lambda_0^2} \right)^{-3}- \frac{1}{2560 }\left( \frac{\tilde u}{\Lambda_0^2} \right)^{-2}+\cdots \right)+\cdots \right],
\end{split}
\end{align}
where $\tilde u :=u-\Lambda_0^2$.
In the following sections, we will generalize these results and compute the quantum corrections to the SW periods
at strong coupling region for the $N_f=1,2,3,4$ cases.

%---------------------------------------------------------------------------------------------------------------------------------------------------------------------------------------------------------------------

\section{Quantum periods for $N_f\geq 1$} 
Let us study the quantum SW periods for $SU(2)$ theory with $N_f\geq 1$ hypermultiplets.  We will choose $N_+$ of (\ref{higgs_moduli}) such that the differential equation (\ref{eq:qswc1}) become the second order differential equation. Then we convert the quantum SW curve into the Schr\"odinger type equation (\ref{eq:qsw1}). The quantum SW periods are given by the integral of (\ref{recursion_relation2}) and (\ref{recursion_relation4}). These periods can be represented as $\hat{\mathcal{O}}_k \Pi ^{(0)}$ with some differential operators $\hat{\cal O}_k$. We will find the second and fourth order corrections to the SW periods.
In the following, $\Delta_{N_f}$ stands for $\Delta$ and $D_{N_f}$ for $D$ in (\ref{general;discriminant}) and (\ref{general;D}) for the $N_f$ theory.
\subsection*{$N_f=1$ theory}
In the theory with $N_f=1$ hypermultiplet, we can take $N_+=1$ in the SW curve (\ref{SW_curve}) without loss of generality. The quantum curve is written as the Schr\"odinger type equation with the 
Tzitz\'eica–Bullough–Dodd type potential:
\begin{align}
Q(x )=-\frac{1}{2} \Lambda_1^{3/2} m_1 e^{i x}-u-\frac{1}{16} \Lambda_1^3 e^{2 i x}-\frac{1}{2} \Lambda_1^{3/2} e^{-i x} ,
\end{align}
where $Q_2(x)=0$. 
The SW periods $\Pi^{(0)}$  satisfy the Picard-Fuchs equation (\ref{eq:pf1}) with 
\begin{align} \label{Nf=1;discriminant}
\begin{split}
\Delta_1=&-\Lambda_1^6( 256u^3-256u^2m_1^2-288um_1\Lambda_1^3+256m_1^3\Lambda_1^3 +27\Lambda_1^6) ,\\
D_{1}=&-16u^2+12m_1 \Lambda_1^3.
\end{split}
\end{align}
It is also found to satisfy the differential equation with respect to the mass parameter $m$:
\begin{align} \label{Nf=1;u-m}
 \frac{\partial^2 \Pi ^{(0)} }{\partial m_1 \partial u} = b_1 \frac{\partial ^2 \Pi ^{(0)}}{\partial u^2} +c_1 \frac{\partial \Pi ^{(0)}}{\partial u},
\end{align}
where 
\begin{align}
b_1=-\frac{16 m_1u-9\Lambda_1^3}{8(4m_1^2-3u)} , \qquad c_1 =-\frac{m_1}{4m_1^2-3u}.
\end{align}
We will calculate the corrections of the second and  fourth orders in $\hbar $ \cite{Huang:2012kn} to the period integrals using (\ref{recursion_relation2}) and (\ref{recursion_relation4}). These corrections are expressed in terms of the basis $\partial_u \Pi^{(0)}$ and $\partial^2_u \Pi^{(0)}$  
\begin{align}
\begin{split} \label{Nf=1periods2;u-diff}
\Pi^{(2)}=&\left(X_2^1 \frac{\partial^2}{\partial u^2} +X_2^2 \frac{\partial }{\partial u} \right) \Pi ^{(0)} 
\end{split} ,\\
\Pi ^{(4)}=&\left(X_4^1 \frac{\partial^2}{\partial u^2} +X_4^2 \frac{\partial }{\partial u} \right) \Pi ^{(0)} ,  \label{Nf=1periods4;u-diff}
\end{align}
where the coefficients in (\ref{Nf=1periods2;u-diff}) are given by 
\begin{align}
\begin{split}
X_2^1=&  -\frac{-9 \Lambda _1^3 m_1-16 m_1^2 u+24 u^2}{48 \left(4 m_1^2-3 u\right)}, \\
X_2^2=&-\frac{3 u-2 m_1^2}{12 \left(4 m_1^2-3 u\right)},
\end{split}
\end{align}
and the coefficients in (\ref{Nf=1periods4;u-diff}) are given by
\begin{align}
\begin{split}
X_4^1 =& \frac{\Lambda_1^{12}}{1440(4m_1^2 -3u) \Delta_1^2}\bigl( -864 \Lambda_1^9 m_1 \left(4350 m_1^2 u+1192 m_1^4+441 u^2\right) \\
&-49152 \Lambda_1^3 m_1 u^2 \left(-455 m_1^2 u^2+609 m_1^4 u-204 m_1^6+267 u^3\right) \\
&+768 \Lambda_1^6 \left(-19593 m_1^2 u^3+42348 m_1^4 u^2-22624 m_1^6 u+6400 m_1^8+8235 u^4\right) \\
&+131072 u^4 \left(15 m_1^2 u^2+6 m_1^4 u-2 m_1^6+9 u^3\right)-729 \Lambda_1^{12} \left(615 u-1792 m_1^2\right)\bigr),
\end{split} \\
\begin{split}
X_4^2=& \frac{\Lambda_1^{12}}{45(4m_1^2 -3u) \Delta_1^2} \bigl(24 \Lambda_1^6 \left(-1080 m_1^2 u^2+4254 m_1^4 u-800 m_1^6+1215 u^3\right) \\
&-768 \Lambda_1^3 m_1 u \left(-185 m_1^2 u^2+267 m_1^4 u-80 m_1^6+159 u^3\right) \\
&+2048 u^3 \left(15 m_1^2 u^2+6 m_1^4 u-2 m_1^6+9 u^3\right)-81 \Lambda_1^9 m_1 \left(235 m_1^2+6 u\right) \bigr).
\end{split}
\displaybreak[1]
\end{align}
 We will compare the quantum prepotential with the NS limit of the Nekrasov partition function in the weak 
coupling region in the next section.
The above representation of the period integrals is suitable to consider the decoupling limit to the pure $SU(2)$ theory, which is defined by $m_1\to \infty$ and $\Lambda_1 \to 0$ with $m_1 \Lambda_1^3=\Lambda _0^4$ being fixed. In the decoupling limit, the second and fourth order corrections (\ref{Nf=1periods2;u-diff}) and (\ref{Nf=1periods4;u-diff}) agree with (\ref{Nf=0;periods2}) and (\ref{eq:nf0pi2pi4}).

In section \ref{sect:strong}, we will study the deformed period integrals in the strong coupling region, where the monopole/dyon becomes massless. 
In this case, the discriminant $\Delta_1$ of the curve has a zero of the first order where the coefficients in  (\ref{Nf=1periods2;u-diff}) and (\ref{Nf=1periods4;u-diff}) become singular. 
Since the SW periods $\Pi^{(0)}$ satisfy the Picard-Fuchs equation (\ref{eq:pf1}) and the differential equation (\ref{Nf=1;u-m}), the differential operator $\hat{\cal O}_k$ in (\ref{eq:hodo1}) for the
higher order corrections is defined modulo such differential operators.
We note that the coefficients of the differential operator for $\Pi^{(2)}$ can be rewritten as
\begin{align} \label{Nf=1;X2coeff} 
X_2^1=\frac{1}{6} u+\frac{1}{6} m_1 b_1, \qquad
X_2^2=\frac{1}{12} +\frac{1}{6} m_1 c_1.
\end{align}
Using the Picard-Fuchs equation (\ref{eq:pf1}) 
and the differential equation (\ref{Nf=1;u-m}), we find that the second order correction to the SW periods
can be expressed as
\begin{align} \label{Nf=1periods2}
\Pi ^{(2)} =&\frac{1}{12} \left( 2u\frac{\partial^2}{\partial u^2} +2m_1 \frac{\partial}{\partial m_1}\frac{\partial }{\partial u} +\frac{\partial }{\partial u} \right) \Pi ^{(0)} .
\end{align}
In the similar way, we find that the fourth order correction to the SW periods
is expressed as
\begin{align}
\begin{split} \label{Nf=1periods4}
\Pi ^{(4)} =& \frac{1}{1440} \biggl( 28 u^2 \frac{\partial^4}{\partial u^4} +124 u \frac{\partial^3}{\partial u^3} +81 \frac{\partial^2}{\partial u^2} \\
&+56 u m_1 \frac{\partial }{\partial m_1}\frac{\partial^3}{\partial u^3} +28m_1^2 \frac{\partial^2}{\partial m_1^2}\frac{\partial^2}{\partial u^2}+132 m_1 \frac{\partial}{\partial m_1} \frac{\partial ^2}{\partial u^2} \biggr) \Pi ^{(0)} .
\end{split}
\end{align}
Since all the coefficients are now regular when $\Delta_1=0$, we can easily calculate the quantum SW periods at the various strong coupling points in the Coulomb branch.

%-------------------------------------------------------------------------------------------------------------------------------------------------------------

\subsection*{$N_f=2$ theory}
In the case of $N_f=2$, we can choose $N_+=1$ or $N_+=2$ in (\ref{higgs_moduli}) for the SW curve (\ref{SW_curve}). The corresponding quantum curves are the second order differential equation in both cases and can be written in the form of the Schr\"odinger type equation but they have apparently different $Q(x)$:
\begin{align}
Q(x)=&-u-\frac{\Lambda_2}{2} \left( m_1 e^{ix} +m_2 e^{-ix} \right)-\frac{\Lambda_2^2}{8} \cos 2x, \quad (N_+=1) \\
Q(x)=&-\frac{e^{ix}\Lambda_2^3+\Lambda_2^2(e^{2ix}(m_1-m_2)^2-2)+8\Lambda_2 e^{ix}(m_1 m_2-u)+16u}{4(-2+e^{ix}\Lambda_2)^2}  \nonumber \\
&+\hbar^2\frac{e^{ix}\Lambda_2}{2(-2+e^{ix}\Lambda_2)^2},  
\quad\quad\quad\quad\quad\quad\quad\quad\quad\quad (N_+=2)
\end{align}
where for the $N_+=2$ case $Q(x)$ includes the $\hbar^2$ term.   
Although the quantum curves look quite different, they are shown to give the same period integrals. One reason is that the SW periods in both cases satisfy the same Picard-Fuchs equation with
the discriminant $\Delta_2$ and $D_2$: 
\begin{align}
\begin{split}
\Delta_2=&\frac{\Lambda_2^{12}}{16}-3 \Lambda_2^{10} m_1 m_2-\Lambda_2^8 \left(8 u^2-36\left( m_1^2+ m_2^2\right) u+27 m_1^4+27 m_2^4+6 m_1^2 m_2^2\right) \\
&+256 \Lambda_2^4 u^2 \left(u-m_1^2\right) \left(u-m_2^2\right)-32 \Lambda _2^6 m_1 m_2 \left(10 u^2-9\left( m_1^2+ m_2^2\right) u+8 m_1^2 m_2^2\right), \\
D_2=&-\frac{3}{4}  \Lambda_2^4+12 \Lambda_2^2 m_1 m_2-16 u^2,
\end{split}
\end{align}
and the differential equations
\begin{align} \label{Nf=2;u-m}
\frac{\partial^2 \Pi^{(0)}}{\partial m_1\partial u}
=\frac{1}{L_2}\left( b_2^{(1)}\frac{\partial^2 \Pi^{(0)}}{\partial u^2}+c_2^{(1)}\frac{\partial \Pi^{(0)}}{\partial u} \right), \\
\frac{\partial^2 \Pi^{(0)}}{\partial m_2\partial u}
=\frac{1}{L_2}\left( b_2^{(2)}\frac{\partial^2 \Pi^{(0)}}{\partial u^2}+c_2^{(2)}\frac{\partial \Pi^{(0)}}{\partial u} \right),
\end{align}
where
\begin{align} 
L_2=&-\Lambda_2^4+8m_1m_2\Lambda_2^2+32\bigl[4m_1^2m_2^2-3u(m_1^2+m_2^2)+2u^2 \bigr],
\nonumber\\
b_2^{(1)}=&3\Lambda_2^4 m_1-4\Lambda_2^2 m_2(3m_1^2-9m_2^2+8u)-64m_2 u(m_1^2-u),  \nonumber\\
c_2^{(1)}=&4 \Lambda_2^2m_2+32m_1(m_2^2-u), \nonumber\\
b_2^{(2)}=&3\Lambda_2^4 m_2-4\Lambda_2^2 m_1(3m_2^2-9m_1^2+8u)-64m_1 u(m_2^2-u) , \nonumber\\
c_2^{(1)}=&4 \Lambda_2^2m_1+32m_2(m_1^2-u).
\label{eq:coeffs1}
\end{align}
Since the SW periods are uniquely determined from the Picard-Fuchs equation with perturbative behaviors
around singularities, the SW periods do not depend on the choice of $N_+$.
We can also check by explicit calculation that the second and fourth order corrections  are
given by   
\begin{align} \label{P22reg}
\Pi ^{(2)}=& \frac{1}{6} \left( 2u\frac{\partial^2}{\partial u^2} +\frac{3}{2} \left( m_1 \frac{\partial}{\partial m_1}\frac{\partial }{\partial u}+m_2 \frac{\partial }{\partial m_2}\frac{\partial }{\partial u} \right) +\frac{\partial }{\partial u} \right) \Pi ^{(0)}, \\ \label{P24reg}
\Pi ^{(4)} =& \frac{1}{360} \biggl[ 28 u^2\frac{\partial^4}{\partial u^4} +120u\frac{\partial^3}{\partial u^3}+ 75\frac{\partial^2}{\partial u^2} \nonumber \\
&+42 \left( u m_1\frac{\partial }{\partial m_1}\frac{\partial^3}{\partial u^3}+u m_2\frac{\partial }{\partial m_2}\frac{\partial^3}{\partial u^3} \right) + \frac{345}{4} \left( m_1 \frac{\partial }{\partial m_1}\frac{\partial ^2}{\partial u^2}+m_2 \frac{\partial }{\partial m_2}\frac{\partial ^2}{\partial u^2} \right)   \nonumber \\
& +\frac{63}{4} \left( m_1^2 \frac{\partial^2}{\partial m_1^2}\frac{\partial^2}{\partial u^2}+m_2^2 \frac{\partial^2}{\partial m_2^2}\frac{\partial^2}{\partial u^2} \right) +\frac{126}{4} m_1m_2 \frac{\partial }{\partial m_1}\frac{\partial }{\partial m_2}\frac{\partial^2}{\partial u^2} \biggr] \Pi ^{(0)} ,
\end{align}
which are independent of $N_+$.
Here we adapt the expression such  that all the coefficients do not have any singularity at singular points in the moduli space.     
Thus we conclude that the quantum SW periods, at least up to the fourth order in $\hbar $,  do not depend on the choice of $N_+$\cite{Zenkevich:2011zx}. 

As explained in the previous sections, the expressions (\ref{P22reg}) and (\ref{P24reg}) are not a unique way to represent the quantum corrections.
With the help of the Picard-Fuchs equation (\ref{eq:pf1})  and the differential equation (\ref{Nf=2;u-m}),  we can rewrite (\ref{P22reg}) in terms of  a basis $\partial_u^2 \Pi ^{(0)}$ and $\partial_u \Pi ^{(0)}$ as
\begin{align} \label{O22u}
\Pi^{(2)}=\left[\left({1\over3}u+{1\over4 L_2}(m_1b_2^{(1)}+m_2 b_2^{(2)}) \right)\frac{\partial^2}{\partial u^2}
+\left({1\over6}+{1\over4 L_2}(m_1c_2^{(1)}+m_2 c_2^{(2)}) \right)
\frac{\partial}{\partial u}\right]\Pi^{(0)},
\end{align} 
where $L_2$, $b_2^{(1)},\cdots c_2^{(2)}$ are given in  (\ref{eq:coeffs1}).
In the decoupling limit where $m_2 \to \infty$ and $\Lambda_2 \to 0$ with $m_2 \Lambda_2^2=\Lambda_1^3$ being fixed, we have the SW periods of the $N_f=1$
theory.
Furthermore, it can be checked that the second and fourth order corrections to the SW periods become those of the $N_f=1$ theory.

%-------------------------------------------------------------------------------------------------------------------------------------------------------------
\subsection*{$N_f=3$ theory}
In the case of $N_f=3$, we can choose $N_+=1$ or $2$ in (\ref{eq:qswc1}). Otherwise, we obtain the third order differential equation. We will take $N_+=2$ without loss of generality. The quantum curve is the Schr\"odinger type equation (\ref{eq:qsw1}) with
\begin{align}
\begin{split}
Q(x)=&\frac{e^{-2 i x} }{16 \left(-2+e^{i x} \Lambda_3^{1/2}\right)^2} \biggl(-4 \Lambda_3-4 e^{3 i x} \Lambda_3^{1/2} \left(m_3 \Lambda_3+8 m_1 m_2-8 u\right)-e^{2 i x} 
\left(\Lambda_3^2-24 m_3 \Lambda_3+64 u\right) \\
&-4 \left(m_1-m_2\right){}^2 e^{4 i x} \Lambda_3+4 e^{i x} \Lambda_3^{1/2} \left(\Lambda_3-8 m_3\right) \biggr)
+\hbar^2 \frac{e^{i x} \Lambda_3^{1/2} }{2 \left(-2+e^{i x} \Lambda_3^{1/2}\right)^2}.
\end{split}
\end{align}

The SW periods satisfy the Picard-Fuchs equation and the differential equations with respect to the mass parameter $m_i$ ($i=1,2,3$) and the moduli parameter $u$. 
Since these equations are rather complicated, we will write down them
 for the  theory with the same mass $m:=m_1=m_2=m_3$. 
 In this case  the discriminant $\Delta_3$ and $D_3$ become
\begin{align}
\Delta_3=& -\frac{\Lambda_3^2 \left(8 m^2+\Lambda_3 m-8 u\right){}^3 \left(256 \Lambda_3 \left(8 m^3-3 m u\right)+8 \Lambda_3^2 \left(3 m^2+u\right)+3 \Lambda_3^3 m-2048 u^2\right)}{4096}, \label{Delta3}\\
D_3=&-\frac{\Lambda_3^4}{256}+12 \Lambda_3 m^3+\Lambda_3^2 \left(u-\frac{9 m^2}{4}\right)-16 u^2. \label{D3}
\end{align}
Then the Picard-Fuchs equation is obtained  by substituting (\ref{Delta3}) and (\ref{D3}) into (\ref{eq:pf1}). 
We can also confirm that  the SW periods satisfy the differential equation:
\begin{align}
\frac{\partial^2 \Pi^{(0)}}{\partial m\partial u}
= b_3\frac{\partial^2 \Pi^{(0)}}{\partial u^2}+c_3\frac{\partial \Pi^{(0)}}{\partial u} \end{align}
where
\begin{align} \label{Nf=3;bc}
b_3=\frac{3 m \left(\Lambda_3 ^2+24\Lambda_3 m-128 u\right)}{16 \left(16 m^2-\Lambda_3 m-4 u\right)} ,\qquad c_3=\frac{12 m}{m \left(\Lambda_3-16 m\right)+4 u}.
\end{align}

We can also calculate the Picard-Fuchs equation for general mass case based on $\Delta_3$ and $D_3$.
In this case we can check that 
the quantum corrections to the SW periods $\Pi^{(0)}$ are expressed as 
\begin{align} 
\begin{split} \label{Nf=3_periods2}
\Pi^{(2)} =&\Biggl[ \left( \frac{5}{6} u -\frac{1}{384} \Lambda_3^2\right) \frac{\partial^2}{\partial u^2}+\frac{1}{2} \sum_{i=1}^{3} m _i \frac{\partial}{\partial m_i} \frac{\partial }{\partial u} + \frac{5}{12} \frac{\partial}{\partial u} \Biggr] \Pi^{(0)} ,
\end{split} \\
\begin{split} \label{Nf=3_periods4}
\Pi^{(4)} =&\Biggl[\frac{7}{10} \left( \frac{5}{6} u -\frac{1}{384} \Lambda_3^2 \right)^2 \frac{\partial^4}{\partial u^4} +\frac{47}{20}\left( \frac{241}{47} \frac{1}{6} u-\frac{1}{384} \Lambda_3^2 \right) \frac{\partial^3}{\partial u^3} + \frac{571}{480} \frac{\partial^2}{\partial u^2} \\
&+ \sum_{i=1}^3 \left(\frac{7}{10} \left(\frac{5}{6} u-\frac{1}{384} \Lambda_3^2 \right) m_i \frac{\partial }{\partial m_i} \frac{\partial^3}{\partial u^3} + \frac{131}{120} m_i \frac{\partial }{\partial m_i} \frac{\partial^2}{\partial u^2} \right)  \\
&+\sum_{i=1}^3 \sum_{j=1}^3 \left( \frac{7}{40} m_i m_j \frac{\partial }{\partial m_i}\frac{\partial }{\partial m_j} \frac{\partial^2}{\partial u^2} \right)  \Biggr] \Pi^{(0)}.
\end{split}
\end{align}
The coefficients  are not singular when $\Delta _3=0$.
With help of the Picard-Fuchs  equation and the differential equation with respect to the mass parameters, we can rewrite the quantum SW periods (\ref{Nf=3_periods2}) and (\ref{Nf=3_periods4}) 
in terms of a basis $\partial_u \Pi ^{(0)}$ and $\partial_u^2 \Pi ^{(0)}$.
For the equal mass case, we  find that 
\begin{align}  \label{Nf=3;periods2u-diff}
\Pi^{(2)} =&\left[ \left( \frac{5}{6} u -\frac{1}{384} \Lambda_3^2+\frac{1}{2} m b_3\right) \frac{\partial^2 }{\partial u^2}+ \left( \frac{5}{12} +\frac{1}{2} m c_3 \right) \frac{\partial }{\partial u} \right] \Pi^{(0)}. 
\end{align}
In this  expression, however, the coefficients become singular at the point  where $\Delta_3=0$.
But this representation is useful to discuss the decoupling limit to the $N_f=0$ theory. 
In the decoupling limit; $m\to \infty $ and $\Lambda_3 \to 0$ with $m^3 \Lambda_3=\Lambda_0^4$ being fixed, the SW periods for $N_f=3$ theory agree with those for the $N_f=0$ theory. 
Moreover, we can show that the second and fourth order corrections to the quantum SW periods 
become those of the $N_f=0$ theory in this limit.

%-------------------------------------------------------------------------------------------------------------------------------------------------------------
\subsection*{$N_f=4$ theory}
In the case of $N_f=4$, we will take $N_+=2$ in (\ref{eq:qswc1}).
Otherwise, we get the third or fourth order differential equation.
The quantum curve can be written in the form of the Schr\"odinger-type equation with
\begin{align}
\begin{split}
Q(x)=&\frac{e^{-2 i x}}{4 \left(-4 \sqrt{q} \cos (x)+q+4\right)^2} \biggl(4 \sqrt{q} e^{3 i x} \left(m_1^2 q+m_2^2 q-m_1 m_2 (q+8)-m_3 m_4 q+8 u\right) \\
&+4 \sqrt{q} e^{i x} \left(m_3^2 q+m_4^2 q-m_3 m_4 (q+8)-m_1 m_2 q+8 u\right) \\
&-e^{2 i x} \left(q \left(\left(m_1^2+m_2^2+m_3^2+m_4^2\right) q -24 \left(m_1 m_2+m_3 m_4\right)\right)+16 (q+4) u\right) \\ 
&-4q e^{4 i x} \left(m_1-m_2\right){}^2 -4q \left(m_3-m_4\right){}^2 \biggr) \\
&+\hbar^2 \frac{\sqrt{q} e^{-i x} \left(q e^{2 i x}-8 \sqrt{q} e^{i x}+q+4 e^{2 i x}+4\right)}{2 \left(-4 \sqrt{q} \cos (x)+q+4\right)^2}.
\end{split}
\end{align}
For simplicity, we consider the case that all the hypermultiplets have the same mass: $m:=m_1=m_2=m_3=m_4$. 
The SW periods $\Pi ^{(0)}$ satisfy the Picard-Fuchs equation (\ref{eq:pf1}) with the discriminant
$\Delta_4$ and $D_4$ which are given by
\begin{align}
\Delta_4=& \frac{2^{24} q^2 \left(m^2-u\right)^4 \left(m^4 (q-16) q+8 m^2 q u+16 u^2\right)}{(q-4)^{10}}, \nonumber\\
D_4=&\frac{16 \left(-m^4 q \left((q-12)^2 q-192\right)-8 m^2 (q-8) q^2 u-16 ((q-4) q+16) u^2\right)}{(q-4)^4}.\label{eq:disc4}
\end{align}
The quantum corrections to the SW periods are expressed in terms of the basis $\partial_u \Pi^{(0)}$ and $\partial^2_u \Pi ^{(0)}$. The second order correction is given by
\begin{align} \label{Nf=4periods2}
\Pi^{(2)} =& \left(X_2^1 \frac{\partial^2}{\partial u^2} +X_2 ^2 \frac{\partial}{\partial u} \right)\Pi^{(0)} ,
\end{align}
where
\begin{align}
\begin{split} \label{Nf=4;X2coeff}
X_2^1 =&-{-18m^4 q+m^4 q^2-8 m^2 u+10 m^2 q u+24 u^2\over 96 m^2}, \\
X_2^2 =& -{-2m^2+m^2 q +6 u\over 48 m^2} .
\end{split}
\end{align}
The fourth order correction is
\begin{align} \label{Nf=4periods4}
\Pi^{(4)}=& \left(X_4^1 \frac{\partial^2}{\partial u^2} +X_4 ^2 \frac{\partial}{\partial u} \right)\Pi^{(0)} ,
\end{align}
where
\begin{align}
\begin{split}
X_4^1=&\frac{1}{ 46080 m^2 \left(m^2-u\right)^2
   \left(m^2 q-4 m^2 \sqrt{q}+4 u\right)^2 \left(m^2 q+4 m^2 \sqrt{q}+4
 u\right)^2} \\
&\times \Bigl(7 m^{14} q^8-399 m^{14} q^7+8484 m^{14} q^6-80616 m^{14} q^5+312480 m^{14}
   q^4-284544 m^{14} q^3\\
&+153600 m^{14} q^2+175 m^{12} q^7 u-7196 m^{12} q^6 u+96504
   m^{12} q^5 u-436320 m^{12} q^4 u
\\
&+266496 m^{12} q^3 u-789504 m^{12} q^2 u+1848
   m^{10} q^6 u^2-51624 m^{10} q^5 u^2+403488 m^{10} q^4 u^2
\\
&-896256 m^{10} q^3
   u^2+2328576 m^{10} q^2 u^2+313344 m^{10} q u^2+10648 m^8 q^5 u^3
\\
&-190176 m^8 q^4
   u^3+820224 m^8 q^3 u^3-1501184 m^8 q^2 u^3-921600 m^8 q u^3+35968 m^6 q^4
   u^4
\\
&-377984 m^6 q^3 u^4+881664 m^6 q^2 u^4-26624 m^6 q u^4-8192 m^6 u^4+70656 m^4
   q^3 u^5
\\
&-344064 m^4 q^2 u^5-325632 m^4 q u^5+24576 m^4 u^5+73728 m^2 q^2 u^6+12288
   m^2 q u^6
\\
&+319488 m^2 u^6+30720 q u^7+122880 u^7\Bigr), \\
X_4^2=& \frac{1}{23040 m^2 \left(m^2-u\right)^2 \left(m^2 q-4 m^2
   \sqrt{q}+4 u\right)^2 \left(m^2 q+4 m^2 \sqrt{q}+4 u\right)^2}\\
   &\times 
\Bigl(7 m^{12} q^7-287 m^{12} q^6+3780 m^{12} q^5-15816 m^{12} q^4+1440 m^{12}
   q^3
-38400 m^{12} q^2\\
&+147 m^{10} q^6 u-4032 m^{10} q^5 u+29736 m^{10} q^4 u-55872
   m^{10} q^3 u+225408 m^{10} q^2 u+30720 m^{10} q u
\\
&
+1260 m^8 q^5 u^2-21768 m^8 q^4
   u^2+88704 m^8 q^3 u^2-221952 m^8 q^2 u^2-133632 m^8 q u^2
\\
&+5608 m^6 q^4 u^3-56768
   m^6 q^3 u^3+147456 m^6 q^2 u^3+7168 m^6 q u^3-2048 m^6 u^3
\\
&+13536 m^4 q^3 u^4-64512
   m^4 q^2 u^4-58368 m^4 q u^4+6144 m^4 u^4+16512 m^2 q^2 u^5+3072 m^2 q
 u^5
\\
&+79872
   m^2 u^5+7680 q u^6+30720 u^6\Bigr).
\end{split}
\end{align}
In the decoupling limit $m\to \infty $ and $q\to 0$ with $m^4 q=\Lambda_0^4$ being fixed, the SW periods coincide with those for the $N_f=0$ theory.
We can also show that
 the second and fourth order corrections of the quantum SW periods (\ref{Nf=4periods2}) and (\ref{Nf=4periods4}) in this limit agree with those for the $N_f=0$ theory .
 
We  can also consider the massless limit, where
the Picard-Fuchs equation 
becomes a simple form:
\begin{align} \label{massless_Nf=4;PF}
\frac{\partial^2 \Pi^{(0)}}{\partial u^2} +\frac{1}{2u} \frac{\partial \Pi^{(0)}}{\partial u} =0.
\end{align}
Note that the coefficients $X^1_k$ and $X_k^2$ in (\ref{Nf=4periods2}) and (\ref{Nf=4periods4}) become singular in the massless limit $m \to 0$. 
In the massless case,  
it is found that 
(\ref{Nf=4periods2}) and (\ref{Nf=4periods4}) are replaced by
\begin{align}
\Pi^{(2)}&=\left( -{u q\over8}\frac{\partial^2}{\partial u^2}+{(q-4)q\over 16u}\frac{\partial}{\partial q}\right) \Pi^{(0)}, \label{massless_Nf=4;periods2}
\\
\Pi^{(4)}&=\left( {-26q+11 q^2\over 2304}\frac{\partial^2}{\partial u^2}-{(q-4)(-52q+35q^2)\over 4608 u^2}\frac{\partial }{\partial q}-{(q-4)^2 q^2\over 288 u^2}\frac{\partial^2}{\partial q^2} \right) \Pi^{(0)}, \label{massless_Nf=4;periods4}
\end{align}
where these formulas include the  derivative with respect to $q$ in addition to
the $u$-derivatives.

In the following sections, we will compute  the quantum SW periods both in the weak and strong coupling regions and compute the deformed (dual) prepotentials.

%-------------------------------------------------------------------------------------------------------------------------------------------------------------
\section{Deformed periods in the weak coupling region}

In this section, 
for the completeness, we will discuss the expansion of the quantum SW periods in the weak coupling region and compute the deformed prepotential for the $N_f$ theories \cite{Huang:2012kn,He:2013fda}. 
Then we  compare the prepotential with the NS limit of  the Nekrasov partition function \cite{Zenkevich:2011zx}.
Note that the deformed prepotentials for $N_f=1,2,4$ are obtained from the classical limit of the conformal blocks of two dimensional conformal field theories \cite{Piatek:2011tp,Ferrari:2012gc,Piatek:2016xhq}.
The SW periods (\ref{SW_periods}) around $u=\infty$ have been given by (\ref{general;a}) and (\ref{general;aD}) \cite{Masuda:1996xj}.
The quantum SW periods can be obtained by acting the differential operators on the SW periods $a^{(0)}$ and $a_D^{(0)}$. 

\subsection{$N_f\leq 3$}
In the case of $N_f=1$, the discriminant $\Delta_1$ and $D_1$ are given by (\ref{Nf=1;discriminant}).
Expanding $a^{(0)}(u)$ and $a^{(0)}_D(u)$ around $u=\infty$ and substituting them into (\ref{Nf=1periods2}) and (\ref{Nf=1periods4}), we  obtain the expansions around $u=\infty$.
They are found to be
\begin{align}
\begin{split}
a(u) =&\sqrt{\frac{u}{2 }}  -\frac{\Lambda_1^3 m_1 \left(\frac{1}{u}\right)^{3/2}}{2^4 \sqrt{2}}+\frac{3 \Lambda_1^6 \left(\frac{1}{u}\right)^{5/2}}{2^{10} \sqrt{2}}+\cdots \\
&+\hbar ^2 \left( -\frac{\Lambda_1^3 m_1 \left(\frac{1}{u}\right)^{5/2}}{2^6 \sqrt{2}}+\frac{15 \Lambda_1^6 \left(\frac{1}{u}\right)^{7/2}}{2^{12} \sqrt{2}} -\frac{35 \Lambda_1^6 m_1^2 \left(\frac{1}{u}\right)^{9/2}}{2^{11} \sqrt{2}}+\cdots \right) \\
&+\hbar ^4 \left( -\frac{\Lambda_1^3 m \left(\frac{1}{u}\right)^{7/2}}{2^8 \sqrt{2}}+\frac{63 \Lambda_1^6 \left(\frac{1}{u}\right)^{9/2}}{2^{14} \sqrt{2}}-\frac{273 \Lambda_1^6 m^2 \left(\frac{1}{u}\right)^{11/2}}{2^{14} \sqrt{2}}+\cdots \right)+\cdots,
\label{eq:periodnf1a}
\end{split}
\end{align}
\begin{align}
\begin{split}
a_D(u)=&-\frac{i}{2\sqrt{2}\pi } \Biggl[ \sqrt{2} a(u)\left( i \pi -3 \log \frac{16u}{\Lambda_1^2}\right) +\left(6 \sqrt{u}+\frac{m_1^2}{\sqrt{u}}+\frac{\frac{m_1^4}{6}-\frac{1}{4} \Lambda_1^3 m_1}{u^{3/2}}+\cdots \right) \\
&+\hbar ^2\left(-\frac{1}{4 \sqrt{u}}-\frac{m_1^2}{12 u^{3/2}}+\frac{-\frac{9}{64} \Lambda_1^3 m_1-\frac{m_1^4}{12}}{u^{5/2}} +\cdots \right) \\
& +\hbar ^4 \left(\frac{1}{160 u^{3/2}}+\frac{7 m_1^2}{240 u^{5/2}}+\frac{\frac{7 m_1^4}{96}-\frac{127 \Lambda_1^3 m_1}{2560}}{u^{7/2}}+\cdots  \right)+\cdots \Biggr].
\end{split}
\end{align}
Solving $u$ in terms of $a$ in (\ref{eq:periodnf1a}) and substituting it into $a_D$, $a_D$ becomes a function of $a$. Then integrating it over $a$, we obtain the deformed prepotential:
\begin{align}
\begin{split}
\mathcal{F}_{1}(a,\hbar) =&\frac{1}{2\pi i} \left[\mathcal{F}^{\text{pert}}_{1}(a,\hbar)+ \sum_{k=0}^{\infty } \sum_{n=1}^{\infty } \hbar^{2k} \mathcal{F}_{1}^{(2k,n)} \left( \frac{1}{a} \right) ^{2n} \right],
\end{split}
\end{align}
where the first few coefficients of $\mathcal{F}^{(2k,n)}_{1}$ ($k=0,1,2$) are listed in the table \ref{table:Nf=1coeffpre}.
\begin{table}[htb]
\begin{align}
\renewcommand{\arraystretch}{1.8}
\begin{array}{|c|c|c|c|c|}\hline
k&\mathcal{F}_{1}^{(2k,1)}&\mathcal{F}_{1}^{(2k,2)}&\mathcal{F}_{1}^{(2k,3)}&\mathcal{F}_{1}^{(2k,4)} \\ \hline
0&\frac{1}{32} \Lambda_1^3 m_1&-\frac{3 \Lambda_1^6}{8192}&\frac{5 \Lambda_1^6 m_1^2}{16384}& -\frac{7 \Lambda_1^9 m_1}{393216} \\
1&0&\frac{1}{256} \Lambda_1^3 m_1&-\frac{15 \Lambda_1^6}{65536}&\frac{21 \Lambda_1^6 m_1^2}{65536}\\
2&0&0 &\frac{\Lambda_1^3 m_1}{2048}&-\frac{63 \Lambda_1^6}{524288}\\
\hline
\end{array} \nonumber
\end{align}
\caption{The coefficients of the prepotential for the $N_f=1$ theory}
\label{table:Nf=1coeffpre}
\end{table}
The perturbative part $\mathcal{F}^{\text{pert}}_{1}(a,\hbar)$ of the prepotential is given by
\begin{align}
\begin{split} \label{Nf=1;pert}
 \mathcal{F}^{\text{pert}}_{1}(a,\hbar)=& -\frac{3}{2}a^2 \log \frac{a^2}{\Lambda_1^2}  +\frac{1}{2}\mathcal{F}^1_s-a^2\log a -\frac{3m_1^2}{4}\\
 &+\hbar^2 \left(-\frac{1}{12} \log a-\frac{1}{96} \frac{\partial^2 \mathcal{F}_s^1}{\partial a^2} +\frac{1}{16} \right)+\hbar^4 \left( -\frac{1}{5760 a^2}+\frac{7}{2^{10}\cdot 3^2 \cdot 5}\frac{\partial ^4\mathcal{F}_s^1}{\partial a^4} \right)+\cdots,
\end{split}
\end{align}
where $\mathcal{F}^1_s$ is defined as \cite{Ohta:1996hq}
\begin{align}
\mathcal{F}^1_s=\left(a+\frac{m_1}{\sqrt{2}}\right)^2 \log \left(a+\frac{m_1}{\sqrt{2}}\right)+\left(a-\frac{m_1}{\sqrt{2}}\right)^2 \log \left(a-\frac{m_1}{\sqrt{2}}\right).
\end{align}

In a similar way, we can calculate the deformed prepotentials for $N_f=2$ and $3$ theories, which are expanded as
\begin{align} \label{Nf=23;prepotential}
\mathcal{F}_{N_f}(a,\hbar) =&\frac{1}{2\pi i} \left[\mathcal{F}^{\text{pert}}_{N_f}(a,\hbar)+ \sum_{k=0}^{\infty } \sum_{n=1}^{\infty } \hbar^{2k} \mathcal{F}_{N_f}^{(2k,n)} \left( \frac{1}{a} \right)^{2n} \right],
\end{align}
where some coefficients $\mathcal{F}_{N_f}^{(2k,n)}$ ($k=0,1,2$) are given in appendix \ref{sec:expansion_pre}.
The perturbative parts are given by
\begin{align}
\begin{split}\label{Nf=2;pert}
\mathcal{F}_{2}^{\text{pert}}(a,\hbar )=& -a^2\log \frac{a^2}{\Lambda_2^2} +\frac{1}{2} \mathcal{F}^2_s-2a^2\log a-\frac{3}{4}(m_1^2+m_2^2)\\
&+\hbar^2 \left(-\frac{1}{12} \log a-\frac{1}{96} \frac{\partial^2 \mathcal{F}_s^2}{\partial a^2} +\frac{1}{8} \right)+\hbar^4 \left( -\frac{1}{5760 a^2}+\frac{7}{2^{10}\cdot 3^2 \cdot 5}\frac{\partial^4\mathcal{F}_s^2}{\partial a^4} \right)+\cdots , 
\end{split}\\
\begin{split}\label{Nf=3;pert}
\mathcal{F}_{3}^{\text{pert}}(a,\hbar )=& -\frac{1}{4} a^2 \log \frac{a^2}{\Lambda_3^2} +\frac{1}{2} \mathcal{F}^3_s-3a^2 \log a-\sum_{i=1}^3 \frac{3}{4}m_i^2 \\
&+\hbar^2 \left(-\frac{1}{12} \log a-\frac{1}{96} \frac{\partial^2 \mathcal{F}_s^3}{\partial a^2} +\frac{3}{16} \right)+\hbar^4 \left( -\frac{1}{5760 a^2}+\frac{7}{2^{10}\cdot 3^2 \cdot 5}\frac{\partial^4\mathcal{F}_s^3}{\partial a^4} \right)+\cdots ,
\end{split}
\end{align}
where $\mathcal{F}^{N_f}_s$ ($N_f=2,3$) is defined  as \cite{Ohta:1996fr}
\begin{align}
\mathcal{F}^{N_f}_s =\sum _{i=1}^{N_f} \left(\left(a+\frac{m_i}{\sqrt{2}}\right)^2 \log \left(a+\frac{m_i}{\sqrt{2}}\right)+\left(a-\frac{m_i}{\sqrt{2}}\right) ^2 \log \left(a-\frac{m_i}{\sqrt{2}}\right)\right).
\end{align}
These deformed prepotentials are shown to be consistent with the decoupling limits. 

We now compare the prepotentials for $N_f=1,2,3$ theories with the NS limit of the Nekrasov partition functions.
By rescaling the parameters $\hbar $, $m_i$ $(i=1,2,3)$, and $\Lambda_{N_f}$ as
\begin{align*}
2\pi i \mathcal{F}(a,\hbar ) \rightarrow \mathcal{F}(a,\epsilon_1) &,&\Lambda_{N_f}  \rightarrow 2^{2/(4-N_f)} \sqrt{2} \Lambda_{N_f} &,& \hbar \rightarrow \sqrt{2} \epsilon_1 &,& m_i \rightarrow \sqrt{2}m_i,
\end{align*}
and then shifting the mass  parameters : $m_i \to m_i+ \epsilon/2$ for a fundamental matter or $m_i \to \epsilon/2-m_i$ for an anti-fundamental matter, 
 we find that the prepotential agrees with that obtained from the Nekrasov partition \cite{Nekrasov:2002qd}. 
 
\subsection{$N_f=4$}
In the case of $N_f=4$, 
after rescaling of the $y$ and $x$ by a factor of $1-\frac{q}{2}$ in the SW curve, we can apply the formulas (\ref{general;a}) and (\ref{general;aD}).
Expanding around $q =0 $ and integrating over $u$, we have the SW periods $a^{(0)}$ and $a_D^{(0)}$  in the weak coupling region.

To simplify the formulas, we consider the equal mass case $m:=m_1=m_2=m_3=m_4$, where the discriminant $\Delta_4$ and $D_4$ are given in (\ref{eq:disc4}).
The deformed prepotential is
\begin{align} \label{Nf=4;prepotential}
\mathcal{F}_{4}=\frac{1}{2\pi i} \left[ \mathcal{F}_{4}^{\text{pert}}(a,\hbar )+\sum_{k=0} ^{\infty } \sum_{n=1}^{\infty } \hbar^{2k} \mathcal{F}_{4}^{(2k,n)} q ^{n}\right],
\end{align}
where the perturbative part is given by
\begin{align}
\begin{split}
\mathcal{F}_{4}^{\text{pert}}(a,\hbar )=&a^2 \log q+\frac{1}{2}\mathcal{F}_{s}^4 -4 a^2 \log a\\
&+\hbar^2 \left( -\frac{1}{12}\log (a)-\frac{1}{96}\frac{\partial^2 \mathcal{F}_s^4}{\partial a^2}  \right) 
+\hbar^4 \left(-\frac{1}{5760a^2}+\frac{7}{2^{10}\cdot 3^2 \cdot 5}\frac{\partial^4 \mathcal{F}_s^4}{\partial a^4}  \right)+\cdots, 
\end{split}
\end{align}
where
\begin{align}
\begin{split}
\mathcal{F}^{4}_s =&4\left( \left(a+\frac{m}{\sqrt{2}}\right)^2 \log \left(a+\frac{m}{\sqrt{2}}\right)+\left(a-\frac{m}{\sqrt{2}}\right)^2 \log \left(a-\frac{m}{\sqrt{2}}\right) \right) .
\end{split}
\end{align}
The first several coefficients $\mathcal{F}_{4}^{(2k,n)}$ for $k=0,1,2$ are given in appendix \ref{sec:Nf=4coeff:pre}.
By rescaling the parameters $\hbar $, $m$ and $q$ as
 \begin{align}
2\pi i \mathcal{F}(a,\hbar )\to \mathcal{F}(a,\epsilon_1)&,& q\to 4 q &,&\hbar \to \sqrt{2} \epsilon_1&,& m \to \sqrt{2} m, 
\end{align}
we find that (\ref{Nf=4;prepotential}) agrees with the prepotential obtained from the NS limit of the Nekrasov partition function of the theory with the equal mass, 
where the mass parameter 
must be shifted as $m_i \to m_i+ \epsilon /2$ for a fundamental matter or $m_i \to \epsilon /2-m_i$ for an anti-fundamental matter ($i=1,\cdots 4$). 

For the massless case $m=0$,  the Picard-Fuchs equation (\ref{massless_Nf=4;PF}) 
has a solution of the form:
\begin{align} \label{masslessNf=4;SWperiods}
\Pi ^{(0)}=f(q) u^{\frac{1}{2}},
\end{align}
where
\begin{align}
f(q)=\frac{\sqrt{2}}{((q-4)q+16)^{1/4}}F\left( \frac{1}{12};\frac{5}{12};1;\frac{108(q-4)^2q^2}{(q^2-4q+16)^3}\right).
\end{align}
Then, using (\ref{massless_Nf=4;periods2}) and (\ref{massless_Nf=4;periods4}), the second and fourth order corrections to the SW periods can be written as
\begin{align}
\Pi ^{(2)}=&\frac{1}{32 \sqrt{u}} \left(q f(q)+2(q-4) \frac{\partial f(q)}{\partial q}\right), \label{masslessNf=4;expansionperiods2} \\
\Pi^{(4)} =&-\frac{q}{9216 u^{3/2}} \left((11 q-26)f(q)+2 (q-4) \left(16 (q-4) q \frac{\partial^2f(q)}{\partial q^2}+(35 q-52) \frac{\partial f(q)}{\partial q}\right)\right). \label{masslessNf=4;expansionperiods4}
\end{align}
It is found that
the prepotential obtained from (\ref{masslessNf=4;SWperiods}), (\ref{masslessNf=4;expansionperiods2}) and (\ref{masslessNf=4;expansionperiods4}) coincides with (\ref{Nf=4;prepotential}) for $m=0$.

\subsection{Deformed effective coupling constant}
From the relation (\ref{general;relation}) and the Picard-Fuchs equation (\ref{eq:pf1}), we can compute the deformed effective coupling.
Differentiating (\ref{general;relation}) with respect to $u$ and applying the Picard-Fuchs equation (\ref{eq:pf1}), we find
\begin{align}
\frac{\partial }{\partial u} \Pi ^{(2k)}= \left( Y_{2k}^1\frac{\partial^2}{\partial u^2}  +Y_{2k}^2 \frac{\partial }{\partial u} \right) \Pi ^{(0)},
\end{align}
where
\begin{align}
Y_{2k}^1:=&-p_1 X_{2k}^1+\frac{\partial X_{2k}^1}{\partial u}+X_{2k}^2,\label{eq:y2k1}\\ 
Y_{2k}^2:=&-p_2 X_{2k}^1+\frac{\partial X_{2k}^2}{\partial u}.
\end{align}
Then taking the $u$-derivative of the quantum SW period $\Pi=\sum_{k=0}^{\infty} \hbar^{2k}\Pi^{(2k)}$, we have 
\begin{align}
\frac{\partial }{\partial u} \Pi=\left( Y_1 \frac{\partial^2}{\partial u^2}+Y_2 \frac{\partial }{\partial u} \right) \Pi^{(0)},
\end{align}
where 
\begin{align}
Y_1=\sum_{n=1}^\infty \hbar^{2n} Y_{2n}^1, \qquad Y_2=1+\sum_{n=1}^\infty \hbar^{2n} Y_{2n}^2.
\end{align}
The deformed effective coupling is defined by
\begin{align}
\tau :=&\frac{\partial_u a_D}{\partial_u a}.
\end{align}
The leading correction to the classical coupling constant 
$
\tau ^{(0)} =\frac{\partial_u a_D^{(0)}}{\partial_u a^{(0)}}
$
is given by
\begin{align}
\tau=\tau^{(0)} \left( 1+\hbar^2 Y_{2}^1 \partial_u \log \tau^{(0)}+ \mathcal{O}(\hbar^4)\right).
\end{align}
Therefore the leading correction to the effective coupling constant is determined by a dimensionless constant $Y_{2}^1$ in
(\ref{eq:y2k1}). 
Also $\partial _u \log \tau ^{(0)}$ is proportional to the beta functions at the weak coupling.

We will evaluate the coefficient $Y_2^{1}$ for some simple cases, where all hypermultiplets have the same
mass $m$.
For $N_f=0$, from the coefficients $X_2^{1}$ and $X_2^{2}$ in (\ref{Nf=0;periods2}) and
$
p_1 =\frac{2u}{u^2-\Lambda^4_0}
$,
one finds 
\begin{align}
Y_2^{1}=& \frac{1}{8}-\frac{u^2}{6 \left(u^2-\Lambda_0^4\right)}.
\end{align}

In a similar way we can compute the coefficient $Y_2^1$ for $N_f\geq 1$.
The results are the followings:
For $N_f=1$, we have
\begin{align}
Y_2^{1}=& \frac{1}{4} +\left( \frac{1}{2} m +\frac{3}{16 }b_1 \right) c_1 -\frac{1}{6}\left(u+m b_1\right) \left(\frac{\partial_u \Delta_1}{\Delta_1}+\frac{3}{4m^2-3u} \right).
\end{align}
For $N_f=2$, 
we have
\begin{align}
\begin{split}
Y_2^{1}=&\frac{1}{2} +\left( \frac{3m}{4}-2 b_2\right) c_2 - \left( \frac{1}{3} u +\frac{m}{4}b_2 \right) \left( \frac{\partial_u \Delta_2}{\Delta_2}-\frac{8 (3m^2-2u)}{8m^2-8u+\Lambda_2^2} \frac{c_2}{m} \right),
\end{split}
\end{align}
where 
\begin{align}
b_2 =\frac{1}{L_2} (b_2^{(1)}+b_2^{(2)}), \qquad
c_2 =\frac{1}{L_2} (c_2^{(1)}+c_2^{(2)}).
\end{align}
For $N_f=3$, we have
\begin{align}
Y_2^1=\frac{5}{4}+\left( \frac{3}{2} m-\frac{1}{6}b_3 \right)-\left(\frac{5}{6} u -\frac{1}{384} \Lambda_3^2+\frac{1}{2} m b_3  \right) \left(\frac{\partial_u \Delta_3}{\Delta_3}-\frac{24 m^2+8u+m\Lambda_3}{-8m^2+8u-m\Lambda_3}\frac{c_3}{m} \right),
\end{align}
where $b_3$ and $c_3 $ is given by (\ref{Nf=3;bc}).
For $N_f=4$,
we find
\begin{align}
Y_2^{1}=&\frac{1-q}{8}-\frac{5 u}{8 m^2} -\frac{1}{96} \left(2 (4-5 q) u-m^2 (q-18) q-\frac{24 u^2}{m^2}\right) \left(\frac{\partial_u\Delta_4}{\Delta_4}+\frac{3}{m^2-u} \right).
\end{align}
We have confirmed that the above formulas are consistent with the decoupling limit and  the  deformed  periods agree with those obtained from the  NS limit of the  Nekrasov partition function explicitly up to the fourth order in $\hbar$.
%----------------------------------------------------------------------------------------------------------
\section{Deformed periods around the massless monopole point}\label{sect:strong}
In this section, we consider the quantum SW periods in the strong coupling region of the theories with $N_f=1,2,3$ hypermultiplets, 
where a  BPS monopole/dyon becomes massless.   
In particular we will consider the point in the $u$-plane such that  the deformed BPS  monopole becomes  massless $a_D(u)=0$.
The dual SW period $a_D^{(0)}$ becomes zero at the massless monopole point
where the discriminant  $\Delta$ of the SW curve and also $z=-27\Delta/4D^3$ become zero.
In the following, we explicitly calculate the expansion of the quantum SW periods around the classical massless monopole point. 
The periods around the dyon massless  point can be analyzed in the same manner.

First we will give some general arguments on the quantum SW periods around the massless monopole point.  
The solution to the Picard-Fuchs equation around the massless monopole point are given by  \cite{Masuda:1996xj}
\begin{align} 
\partial_u a_D^{(0)}&={\sqrt{2}i\over2}(-D)^{-1/4}F\left({1\over12},{5\over12};1; z \right), \label{monogeneralaD} \\
\partial_u a^{(0)}&=\frac{\sqrt{2}}{2} (-D)^{-1/4} \biggl[\frac{3}{2\pi} \ln 12 F\left( \frac{1}{12}, \frac{5}{12}; 1; z \right) -\frac{1}{2\pi } F_*\left( \frac{1}{12}, \frac{5}{12}; 1; z \right) 
\biggr]. \label{monogenerala} 
\end{align}
Let $u_0$ be the massless monopole point in the $u$-plane, where $\Delta$ becomes zero.
In general,  $z$ and $(-D)^{1/4}$  have the following expansion  around $u_0$
\begin{align}
z&=\sum_{n=1}^{\infty}r_n \tilde{u}^n,\qquad (-D)^{-1/4}=\sum_{n=0}^{\infty} s_n \tilde{u}^n, \label{zmono}
\end{align}
where $\tilde{u}=u-u_0$.
Substituting  (\ref{zmono}) into  (\ref{monogeneralaD}) and (\ref{monogenerala}) and integrating with respect to $u$,
the SW periods can be given in the following form
\begin{align}
a^{(0)}_D(\tilde{u})&=\sum_{n=1}^{\infty} B_n \tilde{u}^n , \label{general;aDatmono}\\
a^{(0)}(\tilde{u})&=\frac{i}{2\pi}\left[l a^{(0)}_D(\tilde{u})
\left\{\log(r_l^{1/l}\tilde{u})-\frac{3}{l}\log12 \right\}+\sum_{n=1}^{\infty}A_n \tilde{u}^n
\right], \label{general;aatmono}
\end{align}
where a constant of integration for $a^{(0)}_D$ is fixed by the condition $a^{(0)}_D(0)=0$ and $a^{(0)}(\tilde{u})$ is given up to constant which is independent of $\tilde{u}$. 
The integer $l$ is defined as the smallest integer which gives nonzero $r_n$ i.e. $r_n =0 \;(n < l)$ and $r_l \neq 0$.
$B_n$ and $A_n$ are expressed in terms of $r_n$ and $s_n$. 
First three terms of $B_n$ and $A_n$ are given by
\begin{align}
B_1&=i\frac{s_0}{\sqrt{2}}, \nonumber \\
B_2&=\frac{i}{2\sqrt{2}}\left(s_1+s_0r_1f^{(1)}\right),   \\ 
B_3&=\frac{i}{3\sqrt{2}}\left\{s_2+(s_0r_2+s_1r_1)f^{(1)}+\frac{1}{2}s_0r_1^2f^{(2)}\right\}, \nonumber \\ 
\nonumber \\
A_1&=-l B_1, \nonumber \\  
A_2&=-\frac{l}{2}B_2+\frac{r_{l+1}}{r_l}\frac{1}{2}B_1 
+\frac{i}{2\sqrt{2}}s_0r_1g^{(1)},  \\
A_3&=-\frac{l}{3}B_3+\frac{r_{l+1}}{r_l} \frac{2}{3}B_2+
\left(\frac{r_{l+2}}{r_l}-\frac{r_{l+1}^2}{2r_l^2}\right) \frac{1}{3}B_1 +\frac{i}{3\sqrt{2}}\left\{(s_0r_2+s_1r_1)g^{(1)}+\frac{1}{2}s_0r_1^2g^{(2)}\right\},
\nonumber
\end{align}
where 
\begin{align}
f^{(n)}&=\frac{(1/12)_n(5/12)_n}{n!}, \nonumber \\
g^{(n)}&=\frac{(1/12)_n(5/12)_n}{(n!)^2}\sum_{r=0}^{n-1}\left(\frac{1}{1/12 +r}+\frac{1}{5/12 +r}-\frac{2}{1+r}\right).
\end{align}
The higher order corrections in $\tilde{u}$ can be calculated in a similar way.
Once the SW periods around the massless monopole point are obtained, the quantum SW periods can be calculated by applying the differential operators as is in the weak coupling region. Thus what we have to do is to obtain the explicit value of $u_0$, which is one of the zero of $\Delta$, and the series expansion of $z$ and $(-D)^{1/4}$ around $u_0$. However, for general mass parameters, the expression of $u_0$ is slightly complicated. Therefor we only give explicit expression of the quantum SW periods in simpler cases; massless hypermultiplets and massive hypermultiplets with the same mass. 

Before going to  these examples, we will discuss  an interesting phenomena due to the quantum corrections.
Although the undeformed SW period $a^{(0)}_D(u)$ becomes  zero at the monopole massless point $u=u_0$, 
the deformed SW period $a_D(u)$ is not zero at the same value of $u$. 
This means that the massless monopole point is shifted in the $u$-plane by the quantum correction.
In fact, the quantum SW period $a_D$ around $\tilde u=0$ takes the form $\sum_{k=0}^{\infty}\hbar^{2k}a_D^{(2k)}$ where
\begin{align}
a_D^{(2k)}&=\sum_{n=0}^{\infty}B_n^{(2k)}\tilde{u}^n.
\end{align}
Here $B_n^{(0)}:=B_n$ in (\ref{general;aDatmono}) with $B_0^{(0)}=0$ and $B_1^{(0)}$, $B_0^{(2)}$ and $B_0^{(4)}$ are observed to be non-zero  by explicit calculation.
We then find the massless monopole point $U_0$ of the deformed theory is expressed as
\begin{align} \label{general;U0}
U_0=u_0+\hbar^2 u_1+\hbar^4 u_2 +\cdots,
\end{align}
where $u_1$ and $u_2$ are determined by
\begin{align}
u_1&=-{B_0^{(2)}\over B_1^{(0)}}, \label{eq:u1} \\
 u_2&=-{B_0^{(4)}\over B_1^{(0)}}-{B_1^{(2)}\over B_1^{(0)}}u_1-{B_2^{(0)}\over B_1^{(0)}}u_1^2. \label{eq:u2}
\end{align}
We will compute  these corrections explicitly in the following examples.

\subsection{Massless hypermultiplets}
We discuss the case where mass of the hypermutitplets is zero. This case gives a simple and interesting example since the moduli space admits some discrete symmetry.  We will consider the massless monopole point in the moduli space.
The solution of the Picard-Fuchs equation around the massless monopole point $u_0$ has been studied in 
 \cite{Ito:1995ga}.

\subsection*{$N_f=1$}
For the $N_f=1$ theory, the massless monopole point is $u_0=-3\Lambda_1^2/2^{8/3}$. Around $u_0$ the $z$ and $(-D_1)^{-1/4}$ is 
expanded as
\begin{align}
z&=-\frac{2^{14/3} }{\Lambda_1^2}\tilde{u}-\frac{2^{22/3}\cdot 5 }{3 \Lambda_1^4}
\tilde{u}^2-\frac{47104}{27 \Lambda_1^6}\tilde{u}^3+\cdots,\\
(-D_1)^{-1/4}&=-i \left( \frac{ 2^{1/3}}{3^{1/3} \Lambda_1}
+\frac{2^2}{3^{3/2} \Lambda_1^3}\tilde{u}+\frac{2^{8/3}}{3^{3/2} \Lambda _1^5}\tilde{u}^2+\cdots \right) ,
\end{align}
from which we can read off the coefficients  $r_n$ and $s_n$ in the expansions (\ref{zmono}).

Substituting these coefficients into (\ref{general;aDatmono}) and (\ref{general;aatmono}),
we can obtain the SW periods $(a^{(0)}(u), a_D^{(0)}(u))$. 
Then, using the  relations (\ref{Nf=1periods2}) and (\ref{Nf=1periods4}), we obtain
the expansion of the quantum SW periods around $\tilde u=0$:
\begin{align}
\begin{split}
a_D(\tilde u)=& \left(\frac{\tilde{u}}{2^{1/6}\cdot 3^{1/2} \Lambda_1}+\frac{\tilde{u}^2}{2^{1/2}\cdot 3^{5/2} \Lambda_1^3}+ \frac{\tilde{u}^3}{2^{5/6}\cdot 3^{11/2} \Lambda_1^5}+ \cdots \right) \\ 
&+\frac{\hbar^2 }{\Lambda_1} \left(\frac{5}{2^{19/6} \cdot 3^{5/2} }+ \frac{35}{2^{7/2}\cdot 3^{9/2}}\left( \frac{\tilde{u}}{\Lambda_1^2}\right)+\frac{665 }{2^{23/6}\cdot 3^{15/2}}\left( \frac{\tilde{u}}{\Lambda_1^2}\right)^2 +\cdots \right) \\
&+\frac{\hbar^4}{\Lambda^3_1}\left( \frac{2471}{6^{15/2} }+\frac{144347 }{2^{53/6}\cdot 3^{19/2} }\left( \frac{\tilde{u}}{\Lambda_1^2}\right)+\frac{1964347 }{2^{55/6} \cdot 3^{23/2}}\left( \frac{\tilde{u}}{\Lambda_1^2}\right)^2+\cdots \right)+\cdots ,
\end{split} \label{Nf=1;aDatmono}
\end{align}
\begin{align}
\begin{split}
a(\tilde u)=&\frac{i}{2\pi} \Biggl[ a_D(\tilde{u}) \left( - i\pi +\log \frac{\tilde{u}}{2^{4/3} \ 3^3\ \Lambda_1^2}\right) +i \left(-\frac{ \tilde{u}}{2^{1/6}\cdot 3^{1/2} \Lambda _1}-\frac{5\tilde{u}^2}{2^{3/2
}\cdot 3^{5/2} \Lambda_1^3}-\frac{298 \tilde{u}^3}{2^{5/6}\cdot 3^{13/2} \Lambda_1^5}+\cdots \right) \\
&+\frac{i\hbar^2}{\Lambda_1}\left( -\frac{1}{2^{23/6}\cdot 3^{1/2} }\left( \frac{\tilde{u}}{\Lambda_1^2}\right)^{-1}+\frac{13 }{2^{19/6}\cdot 3^{7/2}}+\frac{101}{6^{9/2}}\left( \frac{\tilde{u}}{\Lambda_1^2}\right)+\cdots\right) \\
&+\frac{i\hbar^4}{\Lambda_1^3} \left(\frac{7 }{2^{15/2}\cdot 3^{1/2}\cdot 5}\left( \frac{\tilde{u}}{\Lambda_1^2}\right)^{-3}+\frac{29}{2^{47/6}\cdot 3^{5/2} \cdot 5 }\left( \frac{\tilde{u}}{\Lambda_1^2}\right)^{-2}+\frac{107 }{2^{49/6}\cdot 3^{9/2}}\left( \frac{\tilde{u}}{\Lambda_1^2}\right)^{-1}+\cdots \right)\Biggr].
\end{split}
\end{align}
Inverting the series of $a_D$ in terms of $\tilde{u}$, we obtain $\tilde{u}$ as a function of $a_D$. Substituting $\tilde{u}$ into $a$ and  integrating $a$ with respect to $a_D$, 
we obtain the dual prepotential:
\begin{align}
\begin{split}
\mathcal{F}_{D1}(a_D,\hbar ) =&\frac{i}{8\pi} \left[a_D^2 \log \left(\frac{a_D}{\Lambda_1} \right) ^2-\frac{\hbar^2}{12}  \log \left(a_D\right)-\frac{7\hbar^4  }{5760 a_D^2}+\cdots \right. \\
&\qquad \qquad+\left. \sum_{k=0}^{\infty} \sum_{n=1}^{\infty } \Lambda_1^2 \left( \frac{\hbar}{\Lambda_1}\right)^{2k}  \mathcal{F}_{D1}^{(2k,n)} \left( \frac{a_D}{\Lambda_1}\right)^n \right],
\end{split}
\end{align}
where the first several coefficients $\mathcal{F}_{D1}^{(2k,n)} $ ($k=0,1,2$) are listed in the table \ref{table:Nf=1coeffdualpre}.
\begin{table}
\begin{align} 
\renewcommand{\arraystretch}{1.8}
\begin{array}{|c|c|c|c|c|} \hline
k&\mathcal{F}_{D1}^{(2k,1)}& \mathcal{F}_{D1}^{(2k,2)}&\mathcal{F}_{D1}^{(2k,3)}&\mathcal{F}_{D1}^{(2k,4)} \\ \hline
0&0&-3&-\frac{5}{12} \frac{1}{\tilde{c}(1)}& -\frac{515}{1152}\frac{1}{\tilde{c}(1)^2}\\
1&\frac{25}{96}\frac{1}{\tilde{c}(1)} &\frac{425}{4608}\frac{1}{\tilde{c}(1)^2} &-\frac{3275}{110592}\frac{1}{\tilde{c}(1)^3} &-\frac{50645}{294912}\frac{1}{\tilde{c}(1)^4} \\
2&\frac{104263}{5308416}\frac{1}{\tilde{c}(1)^3} &\frac{757333}{28311552}\frac{1}{\tilde{c}(1)^4}&-\frac{7173929}{1019215872}\frac{1}{\tilde{c}(1)^5} & -\frac{4749125675}{32614907904}\frac{1}{\tilde{c}(1)^6} \\ 
\hline
\end{array} \nonumber
\end{align}
\caption{The coefficients of the dual prepotetials for the $N_f=1$ theory, where $\tilde{c}(1)=-3^{3/2}\cdot 2^{-17/6}$ \cite{Ito:1995ga}.}
\label{table:Nf=1coeffdualpre}
\end{table}

\subsection*{$N_f=2,3$}
For $N_f=2$, the massless monopole point is  $u_0=\Lambda_2^2/8$. Then $z$ and $(-D_2)^{-1/4}$ are expanded as
\begin{align}
z&= \frac{108 }{\Lambda_2^4}\tilde{u}^2-\frac{432 }{\Lambda_2^6}\tilde{u}^3-\frac{3456 }{\Lambda_2^8}\tilde{u}^4+\cdots ,\\
(-D_2)^{-1/4}&=\frac{1}{\Lambda_2}-\frac{\tilde{u}}{\Lambda_2^3}-\frac{3 \tilde{u}^2}{2 \Lambda_2^5}+\cdots .
\end{align}
Then we have
\begin{align}
\begin{split}
a_D(u)=&i \left(\frac{\tilde{u}}{2^{1/2} \Lambda_2}-\frac{\tilde{u}^2}{2^{3/2} \Lambda_2^3}+\frac{3 \tilde{u}^3}{2^{5/2} \Lambda_2^5}+\cdots \right) \\
&+\frac{i\hbar^2}{\Lambda_2}\left( \frac{1}{2^{7/2} } -\frac{5 }{2^{9/2} }\left( \frac{\tilde{u}}{\Lambda_2^2} \right)+\frac{35 }{2^{11/2} }\left( \frac{\tilde{u}}{\Lambda_2^2} \right)^2+\cdots \right) \\
&+\frac{i\hbar^4}{\Lambda_2^3}\left(-\frac{17}{2^{17/2} }+\frac{721 }{2^{21/2}}\left( \frac{\tilde{u}}{\Lambda_2^2} \right) -\frac{10941 }{2^{23/2} }\left( \frac{\tilde{u}}{\Lambda_2^2} \right)^2+\cdots \right)+\cdots ,
\end{split} \label{Nf=2;aDatmono}
\end{align}
\begin{align}
\begin{split}
a(u)=&\frac{i}{2\pi} \left[ 2a_D(\tilde{u})\log \frac{\tilde{u}}{4\Lambda_2^2} +i \left(-\frac{2 \tilde{u}}{2^{1/2} \Lambda_2}-\frac{3 \tilde{u}^2}{2^{3/2} \Lambda_2^3}+ \frac{12 \tilde{u}^3}{2^{5/2} \Lambda_2^5}+\cdots \right) \right. \\
&+\frac{i\hbar^2}{\Lambda_2}\left( \frac{1}{2^{5/2}\cdot 3}\left( \frac{\tilde{u}}{\Lambda_2^2} \right)^{-1} +\frac{10}{2^{7/2}\cdot 3}-\frac{77 }{2^{9/2}\cdot 3 }\left( \frac{\tilde{u}}{\Lambda_2^2} \right)+\cdots\right) \\
&\left.+\frac{i\hbar^4}{\Lambda_2^3}\left( \frac{7 }{2^{11/2}\cdot 3^2 \cdot 5}\left( \frac{\tilde{u}}{\Lambda_2^2} \right)^{-3}-\frac{1}{2^{13/2}\cdot 5}\left( \frac{\tilde{u}}{\Lambda_2^2} \right)^{-2}+\frac{53}{2^{15/2}\cdot 3\cdot 5 }\left( \frac{\tilde{u}}{\Lambda_2^2} \right)^{-1}+\cdots \right)+\cdots \right] .
\end{split}
\end{align}

For $N_f=3$, the massless monopole point is $u_0=0$. Then $z$ and $(-D_3)^{-1/4}$ are expanded as
\begin{align}
z&=\frac{2^{22}\cdot 3^3 }{\Lambda_3^8}\tilde{u}^4+\frac{2^{31}\cdot 3^3 }{\Lambda _3^{10}}\tilde{u}^5+\frac{2^{34}\cdot 3^5 \cdot 5 }{\Lambda_3^{12}}\tilde{u}^6+\cdots ,\\
(-D_3)^{-1/4}&=\frac{4}{\Lambda_3}+\frac{256 }{\Lambda_3^3}\tilde{u}+\frac{36864 }{\Lambda _3^5}\tilde{u}^2+\cdots .
\end{align}
Then we have
\begin{align}
\begin{split} \label{Nf=3;aDatmono}
a_D(u)=&i\left( \frac{2^{3/2} \tilde{u}}{\Lambda_3}+\frac{2^{13/2} \tilde{u}^2}{\Lambda_3^3}+\frac{2^{11}\cdot 3 \tilde{u}^3}{\Lambda_3^5}+\cdots \right) \\
&+\frac{i\hbar^2}{\Lambda_3} \left(\frac{1}{2^{1/2} }+2^{13/2} \left( \frac{\tilde{u}}{\Lambda_3^2} \right) +2^{19}\cdot 5^2 \left( \frac{\tilde{u}}{\Lambda_3^2} \right)^2 +\cdots \right) \\
&+\frac{i\hbar^4}{\Lambda_3^3} \left( 2^{5/2}\cdot 5 +2^{17/2}\cdot 43 \left( \frac{\tilde{u}}{\Lambda_3^2} \right) + 2^{25/2}\cdot 1141 \left( \frac{\tilde{u}}{\Lambda_3^2} \right)^2+\cdots \right) ,
\end{split}
\end{align}
\begin{align}
\begin{split}
a(u)=&\frac{i}{2\pi} \left[ 4a_D(\tilde{u}) \log \frac{16 \tilde{u}}{\Lambda_3^2} +i \left( -\frac{2^{7/2} \tilde{u}}{\Lambda_3}+\frac{2^{15/2}\cdot 3 \tilde{u}^2}{\Lambda_3^3}+\frac{2^{29/2}\cdot 3 \tilde{u}^3}{\Lambda_3^5}+\cdots \right) \right. \\
&+\frac{i\hbar^2}{\Lambda_3} \left(-\frac{1}{2^{7/2} }\left( \frac{\tilde{u}}{\Lambda_3^2} \right)^{-1}+\frac{2^{7/2} }{3 }+\frac{2^{13/2}\cdot 29 }{3 }\left( \frac{\tilde{u}}{\Lambda_3^2} \right)+\cdots  \right) \\
&+\left. \frac{i \hbar^4 }{\Lambda_3}\left(\frac{7 }{2^{21/2}\cdot 3^2\cdot 5}\left( \frac{\tilde{u}}{\Lambda_3^2} \right)^{-3}-\frac{1}{2^{9/2}\cdot 3\cdot 5 }\left( \frac{\tilde{u}}{\Lambda _3^2} \right)^{-2}+\frac{7}{2^{3/2}\cdot 5 }\left( \frac{\tilde{u}}{\Lambda_3^2} \right)^{-1}+\cdots \right) \right].
\end{split}
\end{align}
We then obtain the deformed dual prepotentials for the $N_f=2$ and $3$ theories,
which are given by
\begin{align} \label{Nf=2;dualpre}
\begin{split}
\mathcal{F}_{D2}(a_D,\hbar )=&\frac{i}{8\pi} \left[2a_D^2 \log \left(\frac{a_D}{\Lambda_2} \right)^2+\frac{ \hbar^2}{6} \log (a_D)-\frac{7 \hbar^4}{2880 a_D^2}+\cdots \right. \\
&\left. \qquad \qquad +\sum_{k=0}^{\infty } \sum _{n=1}^{\infty } \Lambda_2^2 \left( \frac{\hbar}{\Lambda_2}\right)^{2k}  \mathcal{F}_{D2}^{(2k,n)} \left( \frac{a_D}{\Lambda_2}\right)^n \right]
\end{split}
\end{align} 
for $N_f=2$ and 
\begin{align} \label{Nf=3;dualpre}
\begin{split}
\mathcal{F}_{D3}(a_D,\hbar )=&\frac{i}{8\pi} \left[4a_D^2 \log \left(\frac{a_D}{\Lambda_3} \right)^2+\frac{ \hbar^2}{3} \log(a_D)-\frac{7 \hbar^4}{1440 a_D^2}+\cdots \right.\\
&\left. \qquad \qquad +\sum_{k=0}^{\infty} \sum_{n=1}^{\infty } \Lambda_3^2 \left( \frac{\hbar}{\Lambda_3}\right)^{2k}  \mathcal{F}_{D3}^{(2k,n)} \left( \frac{a_D}{\Lambda_3}\right) ^n \right]
\end{split}
\end{align}
for $N_f=3$, 
where the first several coefficients $\mathcal{F}_{DN_f}^{(2k,n)}$ $(N_f=2,3)$ are listed in the table \ref{table:Nf=2coeffdualpre} and the table \ref{table:Nf=3coeffdualpre}.
\begin{table}[htb]
\begin{align} 
\renewcommand{\arraystretch}{1.8}
\begin{array}{|c|c|c|c|c|} \hline
k&\mathcal{F}_{D2}^{(2k,1)}& \mathcal{F}_{D2}^{(2k,2)}&\mathcal{F}_{D2}^{(2k,3)}&\mathcal{F}_{D2}^{(2k,4)} \\ \hline
0&0&-6&\frac{1}{2} \frac{1}{\tilde{c}(2)} &\frac{5}{64} \frac{1}{\tilde{c}(2)^2}\\ 
1&\frac{3}{16}\frac{1}{\tilde{c}(2)} &\frac{17}{256}\frac{1}{\tilde{c}(2)^2} &\frac{205}{6144}\frac{1}{\tilde{c}(2)^3} &\frac{315}{16384}\frac{1}{\tilde{c}(2)^4} \\
2&\frac{135}{32768}\frac{1}{\tilde{c}(2)^3} &\frac{2943}{524288}\frac{1}{\tilde{c}(2)^4} & \frac{69001}{10485760}\frac{1}{\tilde{c}(2)^5} & \frac{1422949}{201326592}\frac{1}{\tilde{c}(2)^6} \\
\hline 
\end{array}\nonumber
\end{align} 
\caption{The coefficients of the dual prepotential for the $N_f=2$ theory, where $\tilde{c}(2)=-i 2^{-5/2} $ \cite{Ito:1995ga}.}
\label{table:Nf=2coeffdualpre}
\end{table}

\begin{table}[htb]
\begin{align} 
\renewcommand{\arraystretch}{1.8}
\begin{array}{|c|c|c|c|c|} \hline
k&\mathcal{F}_{D3}^{(2k,1)}& \mathcal{F}_{D3}^{(2k,2)}&\mathcal{F}_{D3}^{(2k,3)}&\mathcal{F}_{D3}^{(2k,4)} \\ \hline
0&0&-12&  \frac{1}{\tilde{c}(3)}&\frac{5}{32}\frac{1}{\tilde{c}(3)^2}\\ 
1&-\frac{1}{8}\frac{1}{\tilde{c}(3)}&-\frac{5}{128}\frac{1}{\tilde{c}(3)^2} &-\frac{19}{1024}\frac{1}{\tilde{c}(3)^3}& -\frac{85}{8192}\frac{1}{\tilde{c}(3)^4} \\
 2&\frac{37}{49152}\frac{1}{\tilde{c}(3)^3}&\frac{239}{262144}\frac{1}{\tilde{c}(3)^4}&\frac{5221}{5242880}\frac{1}{\tilde{c}(3)^5}&\frac{102949}{100663296}\frac{1}{\tilde{c}(3)^6} \\
\hline 
\end{array}\nonumber
\end{align}
\caption{The coefficients of the dual prepotential for the $N_f=3$ theory, where $\tilde{c}(3)=i 2^{-13/2} $ \cite{Ito:1995ga}.}
\label{table:Nf=3coeffdualpre}
\end{table}
The dual prepotentials include the classical term and one loop term as (\ref{Nf=1;pert}), (\ref{Nf=2;pert}) and (\ref{Nf=3;pert}) in the weak coupling region. These terms also appear in the pure $SU(2)$ theory \cite{He:2010xa}.

Now  we compute the  shifted massless monopole point $U_0$ in the $u$-plane in these examples.
Using the expansion of $a_D$, we obtain
\begin{align}
U_0 =
\left\{ 
\renewcommand{\arraystretch}{1.8}
 \begin{array}{cc}
 \Lambda_0^2-\frac{1}{32} \hbar^2 +\frac{9}{32768 \Lambda_0^2} \hbar^4+\cdots , & N_f=0\\
 -\frac{3\Lambda_1^2}{2^{8/3}} -\frac{5}{72} \hbar^2 -\frac{1571}{2^{22/3}\ 3^7 \Lambda_1^2} \hbar^4+\cdots ,& N_f=1\\
  \frac{\Lambda_2^2}{8} -\frac{1}{8} \hbar^2+\frac{9}{256 \Lambda_2^2} \hbar^4+\cdots ,& N_f=2
 \\
 -\frac{1}{4} \hbar^2 -\frac{4}{\Lambda_3^2} \hbar ^4 +\cdots ,& N_f=3.
  \end{array}
 \right.
\end{align}

In next subsection, we will discuss the expansion around the massless monopole point $u_0$ for the theory with massive hypermultipltes with the same mass. 
%--------------------------------------------------------------------------------------------------------------------------------------------------------------------------------------------------------------------
\subsection{Massive hypermultiplets with the same mass}
We consider the case that all the hypermultiplets have the same mass $m:=m_1=\cdots=m_{N_f}$.
The classical massless monopole point $u_0$ corresponds a solution of the discriminant $\Delta _{N_f}=0$.
 In the $u$-plane, it is found as follows;
\begin{align}
\begin{split}
u_0=&\frac{-64 m^4-216 \Lambda_1^3 m+8 m^2 H_1^{1/3} -H_1^{2/3}}{24 H_1^{1/3}}, \qquad \quad   \text{for } N_f=1,
\end{split} \\
\begin{split}
u_0=&-\frac{\Lambda_2^2}{8}+\Lambda_2 m, \qquad \qquad \qquad \qquad \qquad \qquad \quad \,\, \,\,\, \text{for } N_f=2,
\end{split} \\
\begin{split}
u_0=&\frac{1}{512} \left(\Lambda_3^2-96 \Lambda_3 m+\sqrt{\Lambda_3 \left(\Lambda_3+64 m\right){}^3}\right),
\qquad \text{for } N_f=3
\end{split}
\end{align}
where
\begin{align}
H_1=729 \Lambda_1^6-512 m^6+4320 \Lambda_1^3 m^3+3 \sqrt{3} \left(27 \Lambda_1^4-64 \Lambda_1 m^3\right){}^{3/2}.
\end{align}
In the decoupling limit $m\to \infty$ and $\Lambda_{N_f}\to 0$ with $m^{N_f} \Lambda_{N_f}^{(4-N_f)}=\Lambda_0^4$ being fixed, these points become the massless monopole point $\Lambda_0^2$ of the $N_f=0$ theory.
If we consider the massless limit, these points become the massless monopole points for the massless $N_f$ theory.

We first discuss the $N_f=1$ theory.
Here we consider the  small mass  $|m|\ll \Lambda_1$, where 
$u_0$ is expanded around $m=0$ as \cite{Ohta:1998ib}
\begin{align}
u_0=&-\frac{3 \Lambda_1^2}{2^{8/3}}-\frac{\Lambda_1 m}{2^{1/3}}+\frac{m^2}{3}+\cdots.
\end{align}
From (\ref{general;aDatmono}),
one obtains the expansion of the SW period $a_D^{(0)}$ around $u=u_0$
\begin{align}
\begin{split}
a_D^{(0)}(\tilde u)=& \tilde{u} \left( \frac{ 1}{2^{1/6}\cdot 3^{1/2} \Lambda_1}-\frac{2^{3/2}m^2}{3^{7/2}\Lambda_1^3}+\cdots  \right)+\tilde{u}^2 \left(\frac{1}{2^{1/2}\cdot3^{5/2} \Lambda_1^3}+\frac{ 2^{17/6} m}{3^{7/2} \Lambda_1^4}+\cdots \right) + \cdots ,
\end{split}
\end{align}
where $\tilde{u}=u-u_0$.
By using the relations (\ref{Nf=1periods2}) and (\ref{Nf=1periods4}), we get the quantum SW periods up to the fourth order in $\hbar $ around $u=u_0 $:
\begin{align}
a_D^{(2)}(\tilde u)=&\left( \frac{5 }{2^{13/6}\cdot 3^{5/2} \Lambda_1}-\frac{ m}{2^{5/6}\cdot 3^{7/2} \Lambda_1^2} +\cdots \right) +\tilde{u} \left(\frac{35 }{2^{7/2}\cdot 3^{9/2} \Lambda_1^3}+\frac{5  m}{2^{1/6}\cdot 3^{11/2} \Lambda_1^4}+\cdots \right)+\cdots ,\\
a_D^{(4)}(\tilde u)=&\left( \frac{2471 }{6 ^{15/2} \Lambda_1^3}-\frac{613  m}{2^{31/6}\cdot 3^{15/2} \Lambda_1^4}+\cdots \right)+\tilde{u} \left(\frac{144347 }{2^{53/6}\cdot 3^{19/2} \Lambda_1^5}+\frac{26495  m}{2^{9/2}\cdot 3^{21/2}  \Lambda_1^6}+\cdots \right)+\cdots.
\end{align}
From these expansions, we find that the monopole massless point $U_0$ is 
given by (\ref{general;U0})
where 
\begin{align}
u_0=&-\frac{3 \Lambda_1^2}{2^{8/3}}-\frac{\Lambda_1 m}{2^{1/3}}+\frac{m^2}{3}+\cdots ,\nonumber \\
u_1=&-\frac{5}{2^3\cdot 3^2}+\frac{m}{2^{2/3}\cdot 3^3 \Lambda_1}+ \frac{5 m^2}{2^{1/3}\cdot 3^4 \Lambda _1^2}+\cdots , \nonumber \\
u_2=&-\frac{1571}{2^{22/3}\cdot 3^7 \Lambda_1^2}+\frac{613 m}{2^5\cdot 3^7 \Lambda_1^3}+\frac{11329 m^2}{2^{11/3}\cdot 3^9 \Lambda_1^4}+\cdots .
\end{align}

For $N_f=2$, we find that the massless monopole point $U_0$ is found to be (\ref{general;U0}) where
\begin{align}
u_0=&-\frac{\Lambda_2^2}{8}+\Lambda_2 m,  \nonumber \\
u_1=&-\frac{m-2 \Lambda_2}{32 m-16 \Lambda_2}, \nonumber \\ 
u_2=&\frac{9 \left(-8 \Lambda_2^3+m^3-2 \Lambda_2 m^2-26 \Lambda _2^2 m\right)}{2048 \Lambda_2 \left(\Lambda_2-2 m\right){}^4}.
\end{align}
In the case of $|m|\ll \Lambda _2$, we have
\begin{align}
u_0=&-\frac{\Lambda _2^2}{8}+\Lambda_2 m,  \nonumber \\
u_1=&-\frac{1}{8}-\frac{3 m}{16 \Lambda_2}-\frac{3 m^2}{8 \Lambda_2^2}+\cdots, \nonumber \\ 
u_2=&-\frac{9}{256 \Lambda_2^2}-\frac{405 m}{1024 \Lambda_2^3}-\frac{2385 m^2}{1024 \Lambda_2^4}+\cdots .
\end{align}

For $N_f=3$ with $|m| \ll \Lambda _3$, we have
\begin{align}
u_0=&-\frac{3\Lambda_3 m}{8}- 3m^2+\cdots , \nonumber \\
u_1=&-\frac{1}{4}+\frac{6 m}{\Lambda_3}-\frac{336 m^2}{\Lambda_3^2}+\cdots , \nonumber \\
u_2=&-\frac{4}{\Lambda_3^2}+\frac{888 m}{\Lambda_3^3}-\frac{131904 m^2}{\Lambda_3^4}+\cdots ,
\end{align}
in (\ref{general;U0}).
Note that the first terms in the expansions of $u_1$ and $u_2$ correspond to those in the massless limit.

\begin{figure}[htb]
 \begin{minipage}{0.5\hsize}
  \begin{center}
   \includegraphics[width=70mm]{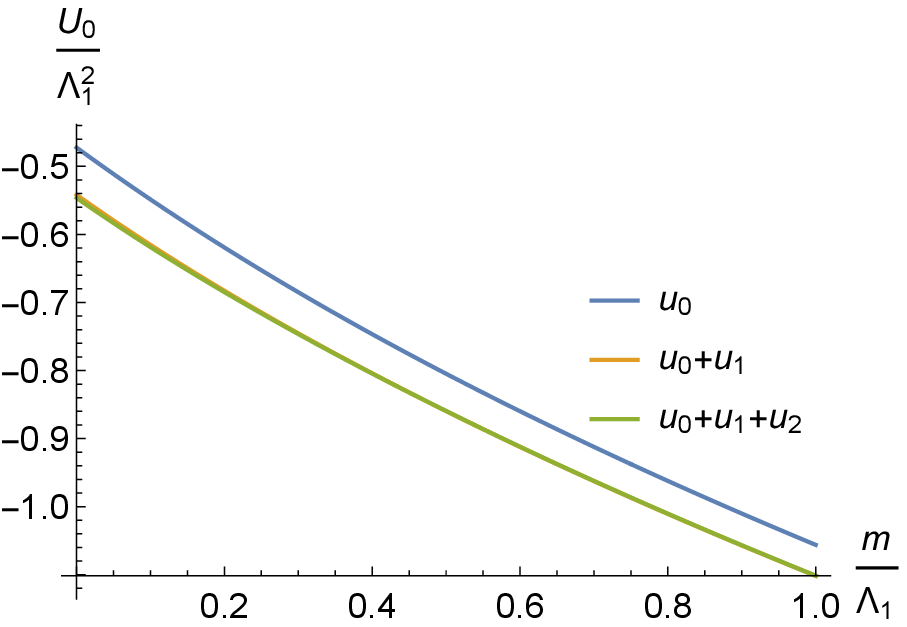}
   \hspace{0.7cm} $N_f=1$
  \end{center}
  \end{minipage}
\begin{minipage}{0.5\hsize}
  \begin{center}
   \includegraphics[width=70mm]{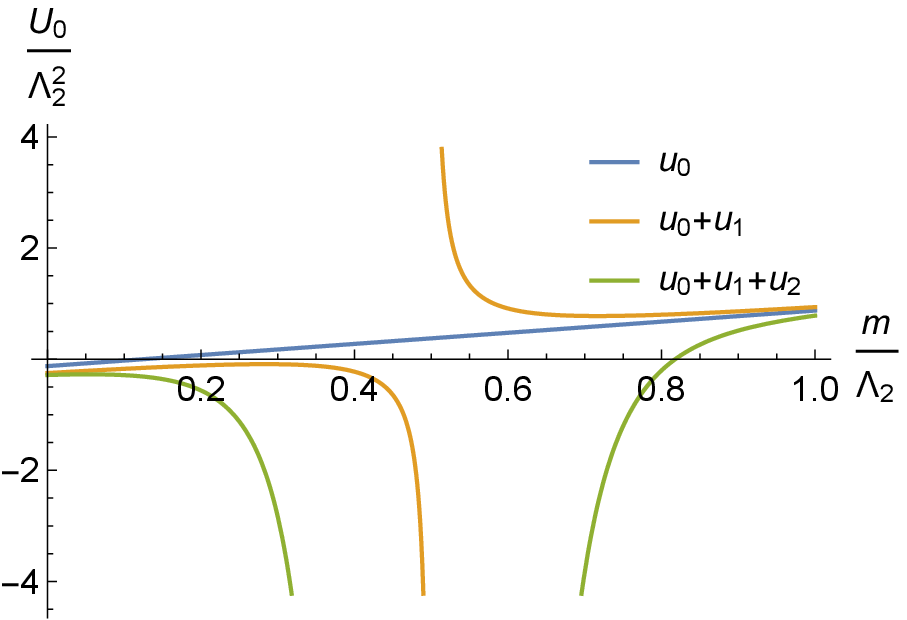}
   \hspace{0.7cm} $N_f=2$
  \end{center}
 \end{minipage}
 \begin{minipage}{0.5\hsize}
  \begin{center}
   \includegraphics[width=70mm]{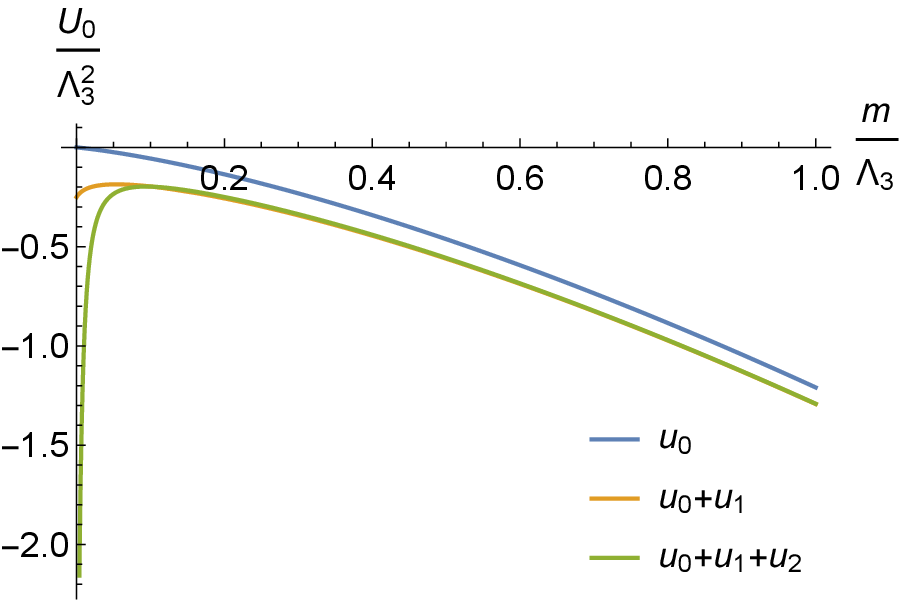}
   \hspace{0.7cm} $N_f=3$
  \end{center}
 \end{minipage}
 \caption{The graphs of $u_0$, $u_0+\hbar^2 u_1$ and $u_0+\hbar^2 u_1+\hbar^4 u_2$
 with respect to $m/\Lambda_{N_f}$ for $N_f=1,2$ and $3$ where we choose $\hbar =1$.}
 \label{fig:one}
\end{figure}

We can perform a similar calculation of $U_0$ up to the fourth order in $\hbar $ for general $m$.
We find that the massless monopole point is shifted by the $\hbar$-correction.
In Fig. \ref{fig:one}
, we have plotted the graphs of the deformed massless monopole point as a function of $m/\Lambda_{N_f}$ where we take $\hbar =1$.
For $N_f=2$,  $U_0$ is singular at the Argyres-Douglas point where $m/\Lambda_2=1/2$. 
This is because the ratios of $B_n^{(k)}$ in (\ref{eq:u1}) and (\ref{eq:u2}) are divergent.
For $N_f=1$ and $3$, however, their ratios are finite.
In order to study the quantum SW periods near the Argyres-Douglas point, we need to rescale the Coulomb moduli and the mass parameters appropriately, which would be 
left for future work.

\section{Conclusions and Discussion}
In this paper, we have studied the low-energy effective theory of ${\cal N}=2$ supersymmetric SU(2)
gauge theory with $N_f$ hypermultiplets 
in the NS limit of the $\Omega$-background.
The deformation of the periods of the SW differential is described by 
the quantum spectral curve, which is 
 the ordinary differential equation and can be solved by the
WKB method.
The quantum spectral curve and 
the Picard-Fuchs equations for the SW periods provide an efficient tool to solve the series expansion with respect to the Coloumb moduli parameter and the deformation parameter $\hbar$. 
We have found  a simple formula to represent the second and fourth order corrections to the SW periods which are obtained by applying some differential operators
acting on the SW periods.
In the weak coupling region we solved the differential equations up to the fourth order in $\hbar$.
We have explicitly checked that the quantum SW periods gives the same prepotential as that obtained from the NS limit of the Nekrasov partition function .

We then studied the quantum corrections expansion around the monopole massless point.
By solving the Picard-Fuchs equations for the SW periods, we have quantum corrections to the dual SW period $a_D$.
We then  found that the monopole massless points in the $u$-plane are shifted by the quantum corrections. 
It is interesting to explore the higher order corrections and how the structure of the  moduli space
is deformed by the quantum corrections.
It is also interesting to study the expansion around the Argyres-Douglas point \cite{Argyres:1995jj,Argyres:1995xn,Eguchi:1996vu,Masuda:1996np} in the $u$-plane where the mutually non-local 
BPS states are massless.
A generalization to the theories with general gauge group and various hypermultiplets is also interesting.

\subsection*{Acknowledgements}
We would like to thank K.~Maruyoshi and K.~Sakai for useful discussions.
 The work of KI is supported
in part by Grant-in-Aid for Scientific Research 
15K05043 and  16F16735 from Japan Society for the Promotion of Science (JSPS).
The work of SK is supported by Iwanami-Fujukai Foundation.
\appendix

\allowdisplaybreaks[1]

\section{$\mathcal{F}_{N_f}^{(2k,n)}$ for the $N_f=2,3$ and $4$ theories} \label{sec:expansion_pre}
In this appendix we explicitly write down some coefficients in the expansion of the prepotentials for $N_f=2,3,4$ theories 
in the weak coupling region.
\subsection{$N_f=2$} \label{sec:Nf=2coeff:pre}
For the $N_f=2$ theory
the first four coefficients of the classical part of the prepotential in (\ref{Nf=23;prepotential}) are 
\begin{align}
\mathcal{F}_{2}^{(0,1)}=&\frac{\Lambda_2^4}{4096}+\frac{1}{32} \Lambda_2^2 m_1 m_2 ,\nonumber \\
\mathcal{F}_{2}^{(0,2)}=&-\frac{3 \Lambda_2^4 m_1^2}{8192}-\frac{3 \Lambda_2^4 m_2^2}{8192},\nonumber \\
\mathcal{F}_{2}^{(0,3)}=&\frac{5 \Lambda_2^8}{134217728}+\frac{5 \Lambda_2^4 m_1^2 m_2^2}{16384}+\frac{5 \Lambda _2^6 m_1 m_2}{196608} , \nonumber \\
\mathcal{F}_{2}^{(0,4)}=&-\frac{63 \Lambda_2^8 m_1^2}{134217728}-\frac{63 \Lambda_2^8 m_2^2}{134217728}-\frac{7 \Lambda_2^6 m_1^3 m_2}{393216}-\frac{7 \Lambda_2^6 m_1 m_2^3}{393216} . 
\end{align}
The coefficients in the second order correction to the prepotential are 
\begin{align}
\mathcal{F}_{2}^{(2,1)}=&0, \nonumber \\
\mathcal{F}_{2}^{(2,2)}=&\frac{\Lambda_2^4}{8192}+\frac{1}{256} \Lambda_2^2 m_1 m_2, \nonumber \\
\mathcal{F}_{2}^{(2,3)}=&-\frac{15 \Lambda_2^4 m_1^2}{65536}-\frac{15 \Lambda_2^4 m_2^2}{65536}, \nonumber\\
\mathcal{F}_{2}^{(2,4)}=&\frac{21 \Lambda_2^8}{134217728}+\frac{21 \Lambda_2^4 m_1^2 m_2^2}{65536}+\frac{35 \Lambda _2^6 m_1 m_2}{786432} .
\end{align}
For the fourth order corrections they are 
\begin{align}
\mathcal{F}_{2}^{(4,1)}=&0,\nonumber\\
\mathcal{F}_{2}^{(4,2)}=&0,\nonumber \\
\mathcal{F}_{2}^{(4,3)}=& \frac{\Lambda_2^4}{16384}+\frac{\Lambda_2^2 m_1 m_2}{2048}, \nonumber\\
\mathcal{F}_{2}^{(4,4)}=&-\frac{63 \Lambda_2^4 m_1^2}{524288}-\frac{63 \Lambda_2^4 m_2^2}{524288}.
\end{align}

\subsection{$N_f=3$}  \label{sec:Nf=3coeff:pre}
For $N_f=3$ the coefficients of the prepotential in the expansion  (\ref{Nf=23;prepotential}) are given by
\begin{align}
\mathcal{F}_{3}^{(0,1)}=&\frac{\Lambda_3^4}{33554432}+\sum_{i=1}^3 \frac{\Lambda_3^2 m_i^2}{4096} +\frac{1}{32} \Lambda_3 m_1 m_2 m_3,\nonumber\\
\mathcal{F}_{3}^{(0,2)}=&\sum_{i=1}^3 -\frac{3 \Lambda_3^4 m_i^2}{33554432} -\sum _{i<j} \frac{3 \Lambda_3^2 m_i^2 m_j^2}{8192}  -\frac{\Lambda_3^3 m_1 m_2 m_3}{32768}, \nonumber\\
\mathcal{F}_{3}^{(0,3)}=&\frac{5 \Lambda_3^8}{4503599627370496}+\sum_{i=1}^3 \left( \frac{5 \Lambda_3^6 m_i^2}{103079215104} +\frac{5 \Lambda_3^4 m_i^4}{134217728}+\frac{5 \Lambda_3^3 m_1 m_2 m_3 m_i^2}{196608} \right)  \nonumber\\
&+\sum _{i<j} \frac{25 \Lambda_3^4 m_i^2 m_j^2}{33554432}  +\frac{5 \Lambda_3^2 m_1^2 m_2^2 m_3^2}{16384}+\frac{7 \Lambda_3^5 m_1 m_2 m_3}{268435456}, \nonumber\\
\mathcal{F}_{3}^{(0,4)}=&\sum_{i=1}^3 \left(-\frac{63 \Lambda_3^8 m_i^2}{2251799813685248}-\frac{7 \Lambda_3^6 m_i^4}{103079215104}-\frac{21 \Lambda_3^5 m_i^2 m_1 m_2 m_3}{268435456}\right) +\sum_{i \neq j}-\frac{63 \Lambda_3^4 m_i^4 m_j^2}{134217728} \nonumber\\
& +\sum_{i<j} \left( -\frac{35 \Lambda_3^6 m_i^2 m_j^2}{34359738368}-\frac{7 \Lambda_3^3 m_i^2 m_j^2 m_1 m_2 m_3}{393216}\right)-\frac{3 \Lambda_3^7 m_1 m_2 m_3}{137438953472}-\frac{147 \Lambda_3^4 m_1^2 m_2^2 m_3^2}{33554432} ,
\end{align}
for the classical part, 
\begin{align}
\mathcal{F}_{3}^{(2,1)}=&-\frac{\Lambda_3^2}{16384}, \nonumber\\
\mathcal{F}_{3}^{(2,2)}=&\frac{5 \Lambda_3^4}{134217728}+\sum _{i=1}^3  \frac{\Lambda_3^2 m_i^2}{8192}+\frac{1}{256} \Lambda_3 m_1 m_2 m_3, \nonumber\\
\mathcal{F}_{3}^{(2,3)}=&-\frac{5 \Lambda_3^6}{412316860416} -\sum_{i=1}^3 \frac{65 \Lambda_3^4 m_i^2}{268435456} -\sum_{i<j}\frac{15 \Lambda_3^2 m_i^2 m_j^2}{65536}-\frac{35 \Lambda_3^3 m_1 m_2 m_3}{786432},\nonumber \\
\mathcal{F}_{3}^{(2,4)}=&\frac{105 \Lambda_3^8}{9007199254740992}+\sum_{i=1}^3 \left( \frac{35 \Lambda_3^6 m_i^2}{103079215104} +\frac{21 \Lambda _3^4 m_i^4}{134217728}+\frac{35 \Lambda_3^3 m_1 m_2 m_3 m_i^2}{786432} \right) \nonumber\\
&+ \sum_{i<j} \frac{147 \Lambda_3^4 m_i^2 m_j^2}{67108864} +\frac{63 \Lambda_3^5 m_1 m_2 m_3}{536870912} +\frac{21 \Lambda_3^2 m_1^2 m_2^2 m_3^2}{65536} ,
\end{align}
for the second order  in  $\hbar $ and
\begin{align}
\mathcal{F}_{3}^{(4,1)}=&0, \nonumber\\
\mathcal{F}_{3}^{(4,2)}=&-\frac{\Lambda_3^2}{32768}, \nonumber\\
\mathcal{F}_{3}^{(4,3)}=&\frac{141 \Lambda_3^4}{2147483648}+\sum_{i=1}^3 \frac{\Lambda_3^2 m_i^2}{16384} +\frac{\Lambda _3 m_1 m_2 m_3}{2048},\nonumber \\
\mathcal{F}_{3}^{(4,4)}=&-\frac{133 \Lambda_3^6}{1649267441664}-\sum_{i=1}^3 \frac{147 \Lambda_3^4 m_i^2}{268435456}-\sum_{i<j}\frac{63 \Lambda_3^2 m_i^2 m_j^2}{524288}-\frac{343 \Lambda_3^3 m_1 m_2 m_3}{6291456},
\end{align}
for the fourth order  in  $\hbar $.

\subsection{$N_f=4$} \label{sec:Nf=4coeff:pre}
For the $N_f=4$ theory
the coefficients of the prepotential (\ref{Nf=4;prepotential})  are given by
\begin{align}
\mathcal{F}_{4}^{(0,1)}=&\frac{a^2}{8}+\frac{m^4}{32 a^2},\nonumber\\
\mathcal{F}_{4}^{(0,2)}=&\frac{13 a^2}{1024}+\frac{11 m^4}{2048 a^2}-\frac{3 m^6}{2048 a^4}+\frac{5 m^8}{16384 a^6},\nonumber\\
\mathcal{F}_{4}^{(0,3)}=&\frac{23 a^2}{12288}+\frac{17 m^4}{16384 a^2}-\frac{m^6}{2048 a^4}+\frac{15 m^8}{65536 a^6}-\frac{7 m^{10}}{98304 a^8}+\frac{3 m^{12}}{262144 a^{10}}, \nonumber \\
\mathcal{F}_{4}^{(0,4)}=&\frac{2701 a^2}{8388608}+\frac{1791 m^4}{8388608 a^2}-\frac{1125 m^6}{8388608 a^4}+\frac{6095 m^8}{67108864 a^6}-\frac{1673 m^{10}}{33554432 a^8} \nonumber \\
&+\frac{2727 m^{12}}{134217728 a^{10}}-\frac{715 m^{14}}{134217728 a^{12}}+\frac{1469 m^{16}}{2147483648 a^{14}},
\end{align}
for the classical part, 
\begin{align}
\mathcal{F}_{4}^{(2,1)}=&\frac{m^4 }{256 a^4},\nonumber\\
\mathcal{F}_{4}^{(2,2)}=&-\frac{m^2}{4096 a^2}+\frac{5 m^4}{4096 a^4}-\frac{15 m^6}{16384 a^6}+\frac{21 m^8}{65536 a^8},\nonumber\\
\mathcal{F}_{4}^{(2,3)}=&-\frac{m^2}{16384 a^2}+\frac{5 m^4}{16384 a^4}-\frac{5 m^6}{12288 a^6}+\frac{91 m^8}{262144 a^8}-\frac{43 m^{10}}{262144 a^{10}}+\frac{55 m^{12}}{1572864 a^{12}}, \nonumber \\
\mathcal{F}_{4}^{(2,4)}=&-\frac{235 m^2}{16777216 a^2}+\frac{2487 m^4}{33554432 a^4}-\frac{8935 m^6}{67108864 a^6}+\frac{11235 m^8}{67108864 a^8}-\frac{38337 m^{10}}{268435456 a^{10}}\nonumber \\
&+\frac{43505 m^{12}}{536870912 a^{12}}-\frac{29549 m^{14}}{1073741824 a^{14}}+\frac{18445 m^{16}}{4294967296 a^{16}},
\end{align}
for the second order in $\hbar$, and
\begin{align}
\mathcal{F}_{4}^{(4,1)}=&\frac{m^4 }{2048 a^6},\nonumber\\
\mathcal{F}_{4}^{(4,2)}=&\frac{1}{65536 a^2}-\frac{m^2}{8192 a^4}+\frac{7 m^4}{16384 a^6}-\frac{63 m^6}{131072 a^8}+\frac{219 m^8}{1048576 a^{10}},\nonumber\\
\mathcal{F}_{4}^{(4,3)}=&\frac{1}{262144 a^2}-\frac{m^2}{32768 a^4}+\frac{119 m^4}{786432 a^6}-\frac{133 m^6}{393216 a^8}+\frac{1689 m^8}{4194304 a^{10}}-\frac{253 m^{10}}{1048576 a^{12}}+\frac{1495 m^{12}}{25165824 a^{14}}, \nonumber \\ 
\mathcal{F}_{4}^{(4,4)}=&\frac{235}{268435456 a^2}-\frac{973 m^2}{134217728 a^4}+\frac{24571 m^4}{536870912 a^6}-\frac{9457 m^6}{67108864 a^8}+\frac{68835 m^8}{268435456 a^{10}}\nonumber \\
&-\frac{625537 m^{10}}{2147483648 a^{12}}+\frac{1765673 m^{12}}{8589934592 a^{14}}-\frac{353325 m^{14}}{4294967296 a^{16}}+\frac{985949 m^{16}}{68719476736 a^{18}},
\end{align}
for the fourth order in $\hbar$.


\begin{thebibliography}{99}
\bibitem{Seiberg:1994rs}
  N.~Seiberg and E.~Witten,
  Nucl.\ Phys.\ B {\bf 426} (1994) 19
   Erratum: [Nucl.\ Phys.\ B {\bf 430} (1994) 485]
  doi:10.1016/0550-3213(94)90124-4, 10.1016/0550-3213(94)00449-8
	[hep-th/9407087].

\bibitem{Seiberg:1994aj}
  N.~Seiberg and E.~Witten,
  %``Monopoles, duality and chiral symmetry breaking in N=2 supersymmetric QCD,''
  Nucl.\ Phys.\ B {\bf 431} (1994) 484
  doi:10.1016/0550-3213(94)90214-3
  [hep-th/9408099].

\bibitem{Argyres:1995jj}
  P.~C.~Argyres and M.~R.~Douglas,
  %``New phenomena in SU(3) supersymmetric gauge theory,''
  Nucl.\ Phys.\ B {\bf 448} (1995) 93
  doi:10.1016/0550-3213(95)00281-V
  [hep-th/9505062].

  
\bibitem{Argyres:1995xn} 
  P.~C.~Argyres, M.~R.~Plesser, N.~Seiberg and E.~Witten,
  %``New N=2 superconformal field theories in four-dimensions,''
  Nucl.\ Phys.\ B {\bf 461}, 71 (1996)
  doi:10.1016/0550-3213(95)00671-0
  [hep-th/9511154].

 
\bibitem{Nekrasov:2002qd}
  N.~A.~Nekrasov,
  %``Seiberg-Witten prepotential from instanton counting,''
  Adv.\ Theor.\ Math.\ Phys.\  {\bf 7} (2004) 831
  [hep-th/0206161].

\bibitem{Nekrasov:2003rj}
  N.~Nekrasov and A.~Okounkov,
  %``Seiberg-Witten theory and random partitions,''
  hep-th/0306238.

\bibitem{Moore:1997dj}
  G.~W.~Moore, N.~Nekrasov and S.~Shatashvili,
  %``Integrating over Higgs branches,''
  Commun.\ Math.\ Phys.\  {\bf 209} (2000) 97
  [hep-th/9712241].



\bibitem{Alday:2009aq}
  L.~F.~Alday, D.~Gaiotto and Y.~Tachikawa,
  %``Liouville Correlation Functions from Four-dimensional Gauge Theories,''
  Lett.\ Math.\ Phys.\  {\bf 91} (2010) 167
  doi:10.1007/s11005-010-0369-5
  [arXiv:0906.3219 [hep-th]].
  
\bibitem{Gaiotto:2009ma} 
  D.~Gaiotto,
  %``Asymptotically free $\mathcal{N} = 2$ theories and irregular conformal blocks,''
  J.\ Phys.\ Conf.\ Ser.\  {\bf 462}, no. 1, 012014 (2013)
  doi:10.1088/1742-6596/462/1/012014
  [arXiv:0908.0307 [hep-th]].
  
  \bibitem{Huang:2009md}
  M.~x.~Huang and A.~Klemm,
  %``Holomorphicity and Modularity in Seiberg-Witten Theories with Matter,''
  JHEP {\bf 1007} (2010) 083
  doi:10.1007/JHEP07(2010)083
  [arXiv:0902.1325 [hep-th]].
  
  \bibitem{Alday:2009fs}
  L.~F.~Alday, D.~Gaiotto, S.~Gukov, Y.~Tachikawa and H.~Verlinde,
  %``Loop and surface operators in N=2 gauge theory and Liouville modular geometry,''
  JHEP {\bf 1001} (2010) 113
  doi:10.1007/JHEP01(2010)113
  [arXiv:0909.0945 [hep-th]].
  
  \bibitem{Maruyoshi:2010iu}
  K.~Maruyoshi and M.~Taki,
  %``Deformed Prepotential, Quantum Integrable System and Liouville Field Theory,''
  Nucl.\ Phys.\ B {\bf 841} (2010) 388
  doi:10.1016/j.nuclphysb.2010.08.008
  [arXiv:1006.4505 [hep-th]].
  
\bibitem{Awata:2010bz} 
  H.~Awata, H.~Fuji, H.~Kanno, M.~Manabe and Y.~Yamada,
  %``Localization with a Surface Operator, Irregular Conformal Blocks and Open Topological String,''
  Adv.\ Theor.\ Math.\ Phys.\  {\bf 16}, no. 3, 725 (2012)
  doi:10.4310/ATMP.2012.v16.n3.a1
  [arXiv:1008.0574 [hep-th]].
  
\bibitem{Nekrasov:2009rc}
  N.~A.~Nekrasov and S.~L.~Shatashvili,
  %``Quantization of Integrable Systems and Four Dimensional Gauge Theories,''
  arXiv:0908.4052 [hep-th].
  
  
  
  \bibitem{Poghossian:2010pn}
  R.~Poghossian,
  %``Deforming SW curve,''
  JHEP {\bf 1104} (2011) 033
  doi:10.1007/JHEP04(2011)033
  [arXiv:1006.4822 [hep-th]].

\bibitem{Mironov:2009uv}
  A.~Mironov and A.~Morozov,
  %``Nekrasov Functions and Exact Bohr-Zommerfeld Integrals,''
  JHEP {\bf 1004} (2010) 040
  doi:10.1007/JHEP04(2010)040
  [arXiv:0910.5670 [hep-th]].
  
     \bibitem{Zenkevich:2011zx}
  Y.~Zenkevich,
  %``Nekrasov prepotential with fundamental matter from the quantum spin chain,''
  Phys.\ Lett.\ B {\bf 701} (2011) 630
  doi:10.1016/j.physletb.2011.06.030
  [arXiv:1103.4843 [math-ph]]. 
  
\bibitem{Beccaria:2016wop} 
  M.~Beccaria,
  %``On the large $\Omega$-deformations in the Nekrasov-Shatashvili limit of $\mathcal N=2^{*}$ SYM,''
  JHEP {\bf 1607}, 055 (2016)
  doi:10.1007/JHEP07(2016)055
  [arXiv:1605.00077 [hep-th]].
  
  %\cite{He:2016khf}
\bibitem{He:2016khf} 
  W.~He,
  %``A new treatment for some periodic Schr\"odinger operators II: the wave function,''
  arXiv:1608.05350 [math-ph].
  

  
  
\bibitem{Mironov:2009dv}
  A.~Mironov and A.~Morozov,
  %``Nekrasov Functions from Exact BS Periods: The Case of SU(N),''
  J.\ Phys.\ A {\bf 43} (2010) 195401
  doi:10.1088/1751-8113/43/19/195401
  [arXiv:0911.2396 [hep-th]].
  
  \bibitem{Popolitov:2010bz}
  A.~Popolitov,
  %``On relation between Nekrasov functions and BS periods in pure SU(N) case,''
  Theor.  Math. Phys. {\bf 178} (2014) 239, 
  arXiv:1001.1407 [hep-th].
  
\bibitem{He:2010xa}
  W.~He and Y.~G.~Miao,
  %``Magnetic expansion of Nekrasov theory: the SU(2) pure gauge theory,''
  Phys.\ Rev.\ D {\bf 82} (2010) 025020
  doi:10.1103/PhysRevD.82.025020
  [arXiv:1006.1214 [hep-th]].
  
\bibitem{Krefl:2014nfa}
  D.~Krefl,
 % ``Non-Perturbative Quantum Geometry II,''
  JHEP {\bf 1412}, 118 (2014),
  doi:10.1007/JHEP12(2014)118
  [arXiv:1410.7116 [hep-th]].

\bibitem{Basar:2015xna} 
  G.~Basar and G.~V.~Dunne,
  %``Resurgence and the Nekrasov-Shatashvili limit: connecting weak and strong coupling in the Mathieu and Lamé systems,''
  JHEP {\bf 1502}, 160 (2015)
  doi:10.1007/JHEP02(2015)160
  [arXiv:1501.05671 [hep-th]].

\bibitem{Kashani-Poor:2015pca} 
  A.~K.~Kashani-Poor and J.~Troost,
  %``Pure $ \mathcal{N}=2 $ super Yang-Mills and exact WKB,''
  JHEP {\bf 1508}, 160 (2015)
  doi:10.1007/JHEP08(2015)160
  [arXiv:1504.08324 [hep-th]].
  
  \bibitem{Ashok:2016yxz}
  S.~K.~Ashok, D.~P.~Jatkar, R.~R.~John, M.~Raman and J.~Troost,
  %``Exact WKB analysis of $ \mathcal{N} $ = 2 gauge theories,''
  JHEP {\bf 1607} (2016) 115
  doi:10.1007/JHEP07(2016)115
  [arXiv:1604.05520 [hep-th]].
  
\bibitem{Basar:2017hpr} 
  G.~Basar, G.~V.~Dunne and M.~Unsal,
  %``Quantum Geometry of Resurgent Perturbative/Nonperturbative Relations,''
  arXiv:1701.06572 [hep-th].
  
    \bibitem{Dorey:1996bn}
  N.~Dorey, V.~V.~Khoze and M.~P.~Mattis,
  %``On N=2 supersymmetric QCD with four flavors,''
  Nucl.\ Phys.\ B {\bf 492} (1997) 607
  doi:10.1016/S0550-3213(97)00132-6
  [hep-th/9611016].  

\bibitem{Hanany:1995na}
  A.~Hanany and Y.~Oz,
  %``On the quantum moduli space of vacua of N=2 supersymmetric SU(N(c)) gauge theories,''
  Nucl.\ Phys.\ B {\bf 452} (1995) 283
  doi:10.1016/0550-3213(95)00376-4
  [hep-th/9505075].

\bibitem{Ceresole:1994fr}
  A.~Ceresole, R.~D'Auria and S.~Ferrara,
  %``On the geometry of moduli space of vacua in N=2 supersymmetric Yang-Mills theory,''
  Phys.\ Lett.\ B {\bf 339} (1994) 71
  doi:10.1016/0370-2693(94)91134-7
  [hep-th/9408036].
  
  
  \bibitem{Klemm:1995wp}
  A.~Klemm, W.~Lerche and S.~Theisen,
  %``Nonperturbative effective actions of N=2 supersymmetric gauge theories,''
  Int.\ J.\ Mod.\ Phys.\ A {\bf 11} (1996) 1929
  doi:10.1142/S0217751X96001000
  [hep-th/9505150].
  
\bibitem{Ito:1995ga} 
  K.~Ito and S.~K.~Yang,
  %``Prepotentials in N=2 SU(2) supersymmetric Yang-Mills theory with massless hypermultiplets,''
  Phys.\ Lett.\ B {\bf 366}, 165 (1996)
  doi:10.1016/0370-2693(95)01310-5
  [hep-th/9507144].
  
  \bibitem{Ohta:1996hq}
  Y.~Ohta,
  %``Prepotential of N=2 SU(2) Yang-Mills gauge theory coupled with a massive matter multiplet,''
  J.\ Math.\ Phys.\  {\bf 37} (1996) 6074
  doi:10.1063/1.531764
  [hep-th/9604051].
  
  \bibitem{Ohta:1996fr}
  Y.~Ohta,
  %``Prepotentials of N=2 SU(2) Yang-Mills theories coupled with massive matter multiplets,''
  J.\ Math.\ Phys.\  {\bf 38} (1997) 682
  doi:10.1063/1.531858
  [hep-th/9604059].
 
\bibitem{Masuda:1996xj} 
  T.~Masuda and H.~Suzuki,
 % ``Periods and prepotential of N=2 SU(2) supersymmetric Yang-Mills theory with massive hypermultiplets,''
  Int.\ J.\ Mod.\ Phys.\ A {\bf 12}, 3413 (1997)
  [Int.\ J.\ Mod.\ Phys.\ A {\bf 12}, 9700179 (1997)]
  doi:10.1142/S0217751X97001791
  [hep-th/9609066].
  
   \bibitem{Erdelyi} A.~Erdelyi et al. , "Higher Transcendental Functions", Vol. 1,
 MacGraw-Hill, New-York 

\bibitem{Huang:2012kn} 
  M.~x.~Huang,
  %``On Gauge Theory and Topological String in Nekrasov-Shatashvili Limit,''
  JHEP {\bf 1206}, 152 (2012)
  doi:10.1007/JHEP06(2012)152
  [arXiv:1205.3652 [hep-th]].
  
  %\cite{He:2013fda}
\bibitem{He:2013fda} 
  W.~He,
  %``N = 2 supersymmetric QCD and elliptic potentials,''
  JHEP {\bf 1411}, 030 (2014)
  doi:10.1007/JHEP11(2014)030
  [arXiv:1306.4590 [hep-th]].

  
\bibitem{Piatek:2011tp} 
  M.~Piatek,
  %``Classical conformal blocks from TBA for the elliptic Calogero-Moser system,''
  JHEP {\bf 1106}, 050 (2011)
  doi:10.1007/JHEP06(2011)050
  [arXiv:1102.5403 [hep-th]].
  
\bibitem{Ferrari:2012gc} 
  F.~Ferrari and M.~Piatek,
  %``Liouville theory, N=2 gauge theories and accessory parameters,''
  JHEP {\bf 1205}, 025 (2012)
  doi:10.1007/JHEP05(2012)025
  [arXiv:1202.2149 [hep-th]].
  
\bibitem{Piatek:2016xhq} 
  M.~Piatek and A.~R.~Pietrykowski,
  %``Classical irregular blocks, Hill’s equation and PT-symmetric periodic complex potentials,''
  JHEP {\bf 1607}, 131 (2016)
  doi:10.1007/JHEP07(2016)131
  [arXiv:1604.03574 [hep-th]].
  

\bibitem{Ohta:1998ib} 
  Y.~Ohta,
  %``Differential equations for scaling relation in N=2 supersymmetric SU(2) Yang-Mills theory coupled with massive hypermultiplet,''
  J.\ Math.\ Phys.\  {\bf 40}, 1891 (1999)
  doi:10.1063/1.532839
  [hep-th/9809180].
  
  \bibitem{Eguchi:1996vu}
  T.~Eguchi, K.~Hori, K.~Ito and S.~K.~Yang,
  %``Study of N=2 superconformal field theories in four-dimensions,''
  Nucl.\ Phys.\ B {\bf 471} (1996) 430
  doi:10.1016/0550-3213(96)00188-5
  [hep-th/9603002].
  
    %\cite{Masuda:1996np}
\bibitem{Masuda:1996np} 
  T.~Masuda and H.~Suzuki,
 % ``On explicit evaluations around the conformal point in N=2 supersymmetric Yang-Mills theories,''
  Nucl.\ Phys.\ B {\bf 495}, 149 (1997)
  doi:10.1016/S0550-3213(97)00199-5
  [hep-th/9612240].
    
 \end{thebibliography}
\end{document}